\documentclass[aps,prx,twocolumn,superscriptaddress,longbibliography]{revtex4-2}

\usepackage{graphicx}
\usepackage{dcolumn}
\usepackage{bm}
\usepackage{amsmath}
\usepackage{amssymb}
\usepackage[braket, qm]{qcircuit}
\usepackage[ruled]{algorithm}
\usepackage{algpseudocode}

\makeatletter
\def\ftype@algorithm{4}
\makeatother
\usepackage{amsmath}
\usepackage{amsfonts}
\usepackage{booktabs}
\usepackage{hyperref}
\hypersetup{
    colorlinks=true,
    linkcolor=blue,
    filecolor=magenta,      
    urlcolor=cyan,
    pdftitle={Overleaf Example},
    pdfpagemode=FullScreen,
    }
\algrenewcommand\algorithmicrequire{\textbf{Input:}}
\algrenewcommand\algorithmicensure{\textbf{Output:}}

\begin{document}

\preprint{APS/123-QED}



\title{\textbf{Ensemble Engineering to Overcome Destructive Cancellation in Quantum Measurements}}

\author{Myeongsu Kim}
 
\affiliation{%
 Department of Computer Science, Purdue University, West Lafayette, Indiana, USA
}%

\author{Manas Sajjan}
\affiliation{
 Department of Electrical and Computer Engineering, North Carolina State University, Raleigh, NC 27695, USA}
\affiliation{National Center for Computational Sciences, Oak Ridge National Laboratory, Oak Ridge, Tennessee 37830, USA}

\author{Sabre Kais}
\altaffiliation{skais@ncsu.edu}
\affiliation{
 Department of Electrical and Computer Engineering, North Carolina State University, Raleigh, NC 27695, USA}
 \affiliation{Department of Chemistry, North Carolina State University, Raleigh, NC 27695, USA
}%

\date{\today}

\begin{abstract}

Quantum state preparation can shape how positive and negative contributions combine in correlation measurements, providing a route to control cancellation through ensemble engineering. We develop a framework for relating this control to the prepared distribution, using the infinite-temperature trace of the same operator product as a uniform-weight reference for ensemble-dependent correlators. For diagonal observables, resolving both projector sectors separates each sector's probability mass from its conditional sign alignment and determines their contributions to the full correlator. A worked example shows what a selected-sector description can miss, and a simple bound limits the error from omitting the complementary contribution.
We use two complementary preparations: Grover-type amplitude amplification to redistribute probability toward a specified target, and oracle-free shallow circuits to shape the distribution at low depth. Zero-time benchmarks combine noiseless simulations with finite-shot sampling and IBM Quantum demonstrations on up to 20 qubits. Sector statistics distinguish low occupation from cancellation within an occupied sector, while comparisons with noiseless simulations show that characteristic features of the engineered cumulative patterns persist on hardware. We also extend the analysis to multi-qubit diagonal observables and connect their sign and sector rules to explicit preparation choices. These results provide a basis for designing and assessing structured ensembles through their operator-resolved contributions to quantum correlation measurements.

\end{abstract}

\maketitle


\section{\label{sec:level1}INTRODUCTION}

Quantum state preparation can shape the weights assigned to basis-resolved operator contributions in correlation measurements. For diagonal observables, uniform weighting can produce exact cancellation through operator orthogonality or symmetry. Random-state methods reproduce the corresponding trace in expectation under suitable ensembles~\cite{approx_trace2,bartsch2009typicality}. Structured preparations provide control over these weights and the resulting ensemble-dependent correlator. Practical implementations must realize these preparations within the circuit-depth and noise constraints of noisy intermediate-scale quantum (NISQ) hardware~\cite{NISQ}.

Our earlier work~\cite{Kim2025ADD} introduced Grover-based and shallow peaked-state constructions and analyzed the projected contribution using support-overlap and biased-expectation diagnostics. For an engineered ensemble, the selected-sector diagnostics leave the complementary contribution undetermined, so identical values of those diagnostics can accompany different full correlators. Evaluating ensemble control thus requires establishing when a selected-sector description is sufficient and what additional information is obtained by resolving both sectors.

The infinite-temperature correlation function (ITCF) serves here as the uniform-weight reference for a fixed operator product. We measure the ensemble-dependent correlator $C_{\mathcal E}(t)$ produced by state preparation, using sector populations and signed sums to distinguish probability transfer from cancellation within each sector. We focus on the short-time, diagonal regime, where the basis contributions can be evaluated from computational-basis measurements. Diagonal Pauli-Z operators and projector-based observables provide an accessible setting for this analysis; related measurement methods have been used in quantum hardware studies of many-body spin dynamics~\cite{superdiffusion_Keerthi}.

We relate preparation choices to the full ensemble-dependent contrast by tracking the probability mass and signed contribution of each sector. A worked example identifies information missing from selected-sector diagnostics, and an elementary bound gives a sufficient condition for a small error from omitting the complementary contribution. For multi-qubit diagonal observables, we make the sign and sector rules explicit and connect specified bit/parity constraints to constructive preparation examples and their logical resource costs. We apply these diagnostics to five-run summaries of 10-qubit Grover and 20-qubit shallow circuits on IBM Quantum hardware, distinguishing low complementary-sector occupation from cancellation within an appreciably occupied sector as origins of a small complementary contribution.

\section{\label{sec:sec2}BACKGROUND: INFINITE-TEMPERATURE CORRELATORS AND ENSEMBLE-INDUCED CANCELLATION}

\subsection{Infinite-Temperature Correlation Function}

For an $n$-qubit system, the (finite-temperature) correlation between two operators $A(t)$ and $B(0)$ may be written as
\begin{equation}
C_{AB}(t) = \mathrm{Tr}\!\left(A(t) B(0)\,\rho(\beta)\right),
\end{equation}
where $A(t)=e^{iHt}A(0)e^{-iHt}$ is the Heisenberg-evolved operator under Hamiltonian $H$, and $\rho(\beta)$ is the thermal state.
In the infinite-temperature limit $\beta\to 0$, the state becomes maximally mixed, $\rho = I/2^n$, yielding the infinite-temperature correlation function (ITCF)
\begin{equation}
C_{\infty}(t) = \frac{1}{2^n}\mathrm{Tr}\!\left(A(t)B(0)\right).
\label{eq:itcf_def}
\end{equation}
Equation~\eqref{eq:itcf_def} is the infinite-temperature reference quantity throughout this work. Its value is fixed by the operator product and can vanish exactly because of operator orthogonality or symmetry. For example, $2^{-n}\mathrm{Tr}(Z_iZ_j)=0$ for Pauli-$Z$ operators on distinct qubits $i\ne j$; this exact zero is the physical reference value. We investigate how the same operator product is averaged when state preparation changes the ensemble. For a distribution $\mathcal E$ over pure states $|r\rangle$, the corresponding \emph{ensemble-dependent correlator} is
\begin{equation}
\label{eq:ensemble_correlator_def}
C_{\mathcal{E}}(t)
:= \mathbb{E}_{|r\rangle\sim \mathcal{E}}
\!\left[\,\langle r|A(t)B(0)|r\rangle\,\right].
\end{equation}
Operationally, this expectation is estimated from measurement outcomes of independently prepared states drawn from $\mathcal E$. Introducing the ensemble-averaged density operator makes its relation to the reference explicit:
\[
\begin{aligned}
\rho_{\mathcal E}
&:=\mathbb{E}_{|r\rangle\sim\mathcal E}[|r\rangle\langle r|],\\
C_{\mathcal E}(t)
&=\mathrm{Tr}\!\left[\rho_{\mathcal E}A(t)B(0)\right].
\end{aligned}
\]
Thus $\rho_{\mathcal E}=I/2^n$ is sufficient for $C_{\mathcal E}(t)=C_\infty(t)$ for every operator product. Haar-random ensembles satisfy this condition. More generally, a particular correlator can agree with the trace even when $\rho_{\mathcal E}\ne I/2^n$; equality is not guaranteed for an arbitrary engineered ensemble.

The computational-basis distribution of the ensemble is
\[
q_z^{(\mathcal E)}
:=\langle z|\rho_{\mathcal E}|z\rangle
=\mathbb E_{|r\rangle\sim\mathcal E}
\!\left[|\langle z|r\rangle|^2\right].
\]
Repeated ideal preparation of a fixed state $|r\rangle$ is the special case $\rho_{\mathcal E}=|r\rangle\langle r|$, for which $q_z^{(\mathcal E)}=|\langle z|r\rangle|^2$.

When $\mathcal{E}$ is unitary-invariant (e.g., Haar-random states), one has
\begin{equation}
\mathbb{E}_{|r\rangle\sim\mathcal{E}}[\langle r|O|r\rangle] = \frac{1}{2^n}\mathrm{Tr}(O),
\end{equation}
so that $C_{\mathcal{E}}(t)$ reduces to the infinite-temperature trace correlator in Eq.~\eqref{eq:itcf_def}.
These relations place the engineered correlator in a state-dependent correlation framework that contains the ITCF as the maximally mixed reference. Changing $\rho_{\mathcal E}$ generally changes the expectation value while leaving the operator product fixed. Throughout this work, cancellation control refers to changing the ensemble weights and their sector contributions in $C_{\mathcal E}(t)$. A larger engineered contrast does not by itself establish a reduction in estimation error or shot cost for the fixed target $C_\infty(t)$.

The connection to the ITCF is used operationally through its basis and sector decomposition: the same operator profile supplies the signed contributions whose weights are controlled by preparation. The measurement task is to determine whether the prepared ensemble changes the contrast through sector probability transfer, intra-sector alignment, or both, and to characterize their realization on hardware. We examine this task at short times (in particular $t\simeq 0$) with diagonal observables, or in settings where diagonal contributions dominate; the zero-time benchmarks use a known operator profile and reference value to make the preparation-dependent contributions directly interpretable.

\subsection{Basis Expansion and Exact Cancellation}

Expanding Eq.~\eqref{eq:itcf_def} in the computational basis $\{\ket{z}\}$ gives
\begin{equation}
C_{\infty}(t)=\frac{1}{2^n}\sum_{z}\bra{z}A(t)B(0)\ket{z}.
\label{eq:itcf_basis}
\end{equation}
For computational-basis diagonal $B(0)$, each trace contribution satisfies the exact relation
\begin{equation}
\bra{z}A(t)B(0)\ket{z}=a_z(t)\,b_z,
\end{equation}
where
\begin{equation}
a_z(t)=\bra{z}A(t)\ket{z},
\qquad
b_z=\bra{z}B(0)\ket{z}.
\end{equation}
Consequently,
\begin{equation}
C_{\infty}(t)=\frac{1}{2^n}\sum_{z} a_z(t)b_z.
\label{eq:itcf_diag}
\end{equation}
This trace identity holds for both diagonal and non-diagonal $A(t)$. In a pure-state expectation, off-diagonal matrix elements can also contribute; the basis-weight description used below is exact when the operator product is diagonal, as in the zero-time benchmarks.

At $t=0$, products of diagonal Pauli strings have basis contributions $f_z:=a_z(0)b_z\in\{+1,-1\}$. For a nonidentity product, the positive and negative contributions balance exactly by Pauli orthogonality. For example, with $D=2^n$,
\begin{equation}
C_\infty(0)=D^{-1}\mathrm{Tr}(Z_iZ_j)=0,\qquad i\ne j.
\label{eq:pauli_trace_zero}
\end{equation}
The operator fixes the sign pattern and the resulting exact cancellation. These analytically known references let us study how preparation changes the weights and the measured correlator. Extending this interpretation to finite time requires examining the actual operator product; its diagonal contributions need not retain equal magnitude.

\subsection{Random-State Fluctuations and Fixed-Trace Sampling}
\label{subsec:trace_sampling}

For general operators, explicitly summing all basis contributions can be costly, and stochastic random-vector or random-state methods provide alternatives \cite{approx_trace1, approx_trace2, richter_simulating}. The zero-time Pauli traces used here are known analytically. To distinguish state fluctuations from measurement noise, consider independent random pure states $\{\ket{r_i}\}_{i=1}^M$ and their exact expectation values:
\begin{equation}
\widetilde{C}_{\infty}(t)=\frac{1}{M}\sum_{i=1}^{M}\bra{r_i}A(t)B(0)\ket{r_i}.
\label{eq:stochastic_trace}
\end{equation}
For Haar-random states,
\begin{equation}
\mathbb{E}_{r}\!\left[\bra{r}O\ket{r}\right]=\frac{1}{2^n}\mathrm{Tr}(O),
\end{equation}
which underlies the trace representation used in dynamical typicality \cite{bartsch2009typicality, elsayed2013regression}. Writing $\ket{r_i}=\sum_z c_{z,i}\ket{z}$, a diagonal operator product gives
\begin{equation}
\bra{r_i}A(0)B(0)\ket{r_i}=\sum_z |c_{z,i}|^2\,f_z.
\label{eq:weighted_sum_general}
\end{equation}
The Haar mean of each basis weight is $1/D$. For a Hermitian $O$, let $X_r=\langle r|O|r\rangle$ and $\mu=\mathrm{Tr}(O)/D$. The Haar second moment gives
\[
\operatorname{Var}_{\rm Haar}(X_r)
=\frac{\mathrm{Tr}(O^2)/D-\mu^2}{D+1}.
\]
For a nonidentity Pauli product, $O^2=I$ and $\mu=0$, so the standard deviation of the exact pure-state expectation is $1/\sqrt{D+1}=O(2^{-n/2})$. Averaging $M$ independent exact expectations divides this variance by $M$. This concentration around the trace is the typicality mechanism supporting random-state methods, including pure-state propagation approaches \cite{steinigeweg2014spin}.

Finite-shot estimation is a separate operation. For $N$ independent computational-basis samples from the uniform distribution, the outcomes $f_z=\pm1$ have mean $\mu$ and the sample mean has variance $(1-\mu^2)/N$. Thus the exact trace, the fluctuation of exact expectations across Haar states, and the sampling error of measured outcomes are distinct quantities. Hardware noise can additionally change the prepared distribution or recorded outcomes.

For the same fixed target $\mu=D^{-1}\sum_z f_z$, consider independent basis samples $z_\ell\sim q$, with $q_z>0$ for every basis state. The ordinary unbiased importance-sampling estimator is
\[
\widehat{\mu}_q=\frac{1}{N}\sum_{\ell=1}^{N}\frac{f_{z_\ell}}{Dq_{z_\ell}},
\qquad \mathbb{E}_q[\widehat{\mu}_q]=\mu.
\]
Since $f_z^2=1$, its variance is
\[
\begin{aligned}
\operatorname{Var}_q(\widehat{\mu}_q)
&=\frac{1}{N}\left[\frac{1}{D^2}\sum_z\frac{1}{q_z}-\mu^2\right]\\
&\ge\frac{1-\mu^2}{N}.
\end{aligned}
\]
Cauchy--Schwarz gives $\sum_z1/q_z\ge D^2$, with equality at $q_z=1/D$. Uniform sampling therefore minimizes variance within this estimator class for the equal-magnitude Pauli contributions. This minimum applies to independent importance samples with the displayed weights. Other measurement protocols and uses of known symmetry require their own optimality analysis.

An unweighted mean of samples from the engineered distribution instead has expectation $C_{\mathcal E}(0)=\sum_z q_z f_z$. The circuits below realize and characterize this ensemble-dependent quantity. The zero-time benchmarks use known sign structure to connect state preparation to sector mass and alignment. Their purpose is to resolve these control mechanisms and their hardware implementation. General finite-time operators require a separate variance analysis.

\subsection{Projector-Based Reformulation and the Key Structural Form}

To expose the cancellation structure in its simplest form, we first adopt a projector decomposition for a single-qubit Pauli-$Z$ observable acting on qubit $k$,
\begin{equation}
Z_k = 2P_{\uparrow}-I, 
\qquad 
P_{\uparrow}\equiv |0\rangle\langle 0|_k\otimes I_{\bar{k}},
\qquad 
P_{\downarrow}\equiv I-P_{\uparrow}.
\label{eq:projector_decomp}
\end{equation}
Substituting $B(0)=Z_k$ into the finite-sample estimator gives the exact algebraic form
\begin{equation}
\widetilde{C}_{AB}(t)
=
\frac{1}{M}\sum_{i=1}^{M}
\left(
2\bra{r_i}A(t)P_{\uparrow}\ket{r_i}
-
\bra{r_i}A(t)\ket{r_i}
\right).
\label{eq:proj_estimator_exact}
\end{equation}
The first term can be decomposed into sector-diagonal and off-sector contributions by inserting
$I=P_{\uparrow}+P_{\downarrow}$ to the left of $A(t)P_{\uparrow}$:
\begin{equation}
A(t)P_{\uparrow}
=
P_{\uparrow}A(t)P_{\uparrow}
+
P_{\downarrow}A(t)P_{\uparrow}.
\label{eq:sector_decomp}
\end{equation}
Thus,
\begin{align}
\widetilde{C}_{AB}(t)
&=
\frac{1}{M}\sum_{i=1}^{M}
\Bigl(
2\bra{r_i}P_{\uparrow}A(t)P_{\uparrow}\ket{r_i}
- \bra{r_i}A(t)\ket{r_i}
\notag\\
&\qquad
+ 2\bra{r_i}P_{\downarrow}A(t)P_{\uparrow}\ket{r_i}
\Bigr).
\label{eq:proj_estimator_decomp}
\end{align}
In the short-time diagonal regime considered below, $A(t)$ is approximately sector diagonal with respect to the $P_{\uparrow}/P_{\downarrow}$ decomposition. Equivalently, the off-sector contribution 
$\bra{r_i}P_{\downarrow}A(t)P_{\uparrow}\ket{r_i}$ is negligible. Retaining only the sector-diagonal contribution gives
\begin{equation}
\widetilde{C}_{AB}(t)\approx 
\frac{1}{M}\sum_{i=1}^{M}
\left(
2\bra{r_i}P_{\uparrow}A(t)P_{\uparrow}\ket{r_i}
-
\bra{r_i}A(t)\ket{r_i}
\right).
\label{eq:proj_estimator_M}
\end{equation}
For intuition, it is convenient to view this sector-diagonal finite-sample expression through a representative prepared state $\ket{r}$, giving the schematic form
\begin{equation}
\widetilde{C}_{AB}(t)\approx
2\bra{r}P_{\uparrow}A(t)P_{\uparrow}\ket{r}
-
\bra{r}A(t)\ket{r}.
\label{eq:proj_estimator_single}
\end{equation}
Equations~\eqref{eq:proj_estimator_M} and \eqref{eq:proj_estimator_single} are sector-diagonal reductions of the exact expression in Eq.~\eqref{eq:proj_estimator_decomp}. Their applicability at finite time requires the off-sector contribution to be negligible. Equation~\eqref{eq:proj_estimator_single} supplies a heuristic representative-state picture; the operational object is the ensemble or finite-sample average in Eq.~\eqref{eq:proj_estimator_M}.

Using $A(t)=V^{\dagger}(t)A(0)V(t)$ with $V(t)=e^{-iHt}$ and inserting computational-basis resolutions of identity into the first term yields
\begin{align}
\langle r|P_{\uparrow}A(t)P_{\uparrow}|r\rangle
&= \sum_{j,k}
\langle r|P_{\uparrow}V^{\dagger}(t)|z_j\rangle
\langle z_j|A(0)|z_k\rangle \notag\\
&\qquad
\langle z_k|V(t)P_{\uparrow}|r\rangle .
\end{align}

If (i) $A(0)$ is diagonal in the computational basis and (ii) we restrict to the short-time regime where $V(t)\approx I$ (in particular $t\to 0$), then the $j\neq k$ contributions vanish and the expression simplifies to a weighted sum over basis components:
\begin{equation}
\bra{r}P_{\uparrow}A(t)P_{\uparrow}\ket{r}
\approx
\sum_{j}\left|\bra{r}P_{\uparrow}\ket{z_j}\right|^2\bra{z_j}A(0)\ket{z_j}.
\label{eq:weighted_sum_pz_az}
\end{equation}
In the same limit, $\bra{r}A(t)\ket{r}\approx \bra{r}A(0)\ket{r}$.
Combining these gives the compact structural form
\begin{align}
\widetilde{C}_{AB}(0)
&\approx
\sum_{j}
2\left|\bra{r}P_{\uparrow}\ket{z_j}\right|^2
\bra{z_j}A(0)\ket{z_j} \notag\\
&\qquad-\bra{r}A(0)\ket{r}.
\label{eq:final_structural_form}
\end{align}
Equation~\eqref{eq:final_structural_form} provides the central structural representation of this work. 

This form resolves the observable into a weighted sum over computational-basis contributions. In this representation, the weights 
\begin{equation}
p_j=\left|\bra{r}P_{\uparrow}\ket{z_j}\right|^2,\qquad 
a_j=\bra{z_j}A(0)\ket{z_j},
\end{equation}
are directly induced by the state-preparation ensemble, while the signs and magnitudes are determined by the operator profile. For the sign-balanced profiles considered here, mean-uniform weights \(\{p_j\}\) average the operator sign structure \(\{a_j\}\) to produce structural cancellation. Finite-shot fluctuations describe the sampling uncertainty around the resulting expectation.

We use state preparation to control the weights $\{p_j\}$ relative to the fixed operator profile. Concentrating weight in regions with aligned contributions changes the ensemble-dependent correlator and exposes their operator-resolved structure.

\section{\label{sec:sec3}Structural diagnostics of cancellation and its breakdown}

This section uses the basis- and sector-resolved representation of Sec.~II to diagnose cancellation in computational-basis contributions. Separating ensemble weights from the operator profile lets us identify how probability transfer and signed alignment shape the measured correlator.

In the state-resolved formulas below, we first fix the prepared state $|r\rangle$, so $C_{\mathcal E}(0)$ refers to the one-state ensemble $\mathcal E=\{|r\rangle\}$. For a general preparation ensemble, the same diagonal formulas apply with $q_z=q_z^{(\mathcal E)}$ and $p_z=\delta_{z_k,0}q_z^{(\mathcal E)}$, and with each state expectation replaced by its ensemble average. Conditional sector means are formed from the resulting ensemble masses and signed sums. The Haar visualizations below show individual sampled states.

\subsection{\label{sec:haar_baseline}Haar-random ensembles: unstructured weighting and destructive cancellation (baseline)}

Our starting point is the structural form derived in Sec.~\ref{sec:sec2}. 
At $t=0$, for diagonal $A(0)$ and $B(0)=Z_k$, the fixed-state correlator is a weighted sum over basis contributions, where the weights are determined by the prepared state. Under Haar-random or otherwise unstructured ensembles, these weights, while fluctuating across individual basis states, exhibit no systematic alignment with the operator profile. As a result, the structural decomposition predicts extensive cancellation whenever the operator profile exhibits sign-alternating behavior.
\begin{equation}
\label{eq:Ce_shorttime}
C_{\mathcal{E}}(0) = 2 \sum_{z} p_z a_z(0) - \langle r|A(0)|r\rangle,
\end{equation}
where
\begin{equation}
a_z(0)\equiv \langle z|A(0)|z\rangle,\qquad
p_z \equiv \big|\langle r|P_{\uparrow}|z\rangle\big|^2.
\end{equation}
Here $a_z(0)$ is the diagonal operator profile, while $p_z$ is the projector-induced weight associated with the prepared state.
Equation~\eqref{eq:Ce_shorttime} makes explicit that the observable is governed by the interplay between the basis-resolved operator structure and the ensemble-induced weighting.

For visualization, we order computational-basis states as $\{|z_i\rangle\}_{i=0}^{2^n-1}$ by their integer index and consider both the signed contributions $p_{z_i}a_{z_i}(0)$ and their cumulative sum
\begin{equation}
S_\uparrow(i)\equiv \sum_{j=0}^{i} p_{z_j}\,a_{z_j}(0).
\label{eq:cumsum_def}
\end{equation}
Persistent growth of $S_\uparrow(i)$ indicates coherent addition of contributions within the selected sector, while erratic fluctuations around zero indicate extensive destructive cancellation. Its endpoint is the selected-sector signed sum $W_\uparrow$ defined below.

\subsubsection{Mean-uniform, structureless weighting under Haar sampling}

Haar-random states have mean-uniform weights with no systematic alignment to a fixed operator profile. Sufficiently deep local random quantum circuits can approximate the corresponding Haar moments by forming approximate unitary designs~\cite{approx_poly}. For a Haar-random state in dimension \(d=2^n\), the computational-basis probabilities
\(q_z=|\langle z|r\rangle|^2\) satisfy
\[
\mathbb{E}_{\rm Haar}[q_z]=\frac{1}{d},
\qquad
{\rm Var}_{\rm Haar}(q_z)=\frac{d-1}{d^2(d+1)} ,
\]
so individual basis weights fluctuate at the same order as their mean. For the projected weights used here, \(p_z=\delta_{z_k,0}q_z\), the typical nonzero scale on the selected sector is therefore \(O(2^{-n})\), but the fluctuations carry no systematic dependence on \(z\) and no systematic correlation with \(a_z(0)\). Haar-like weighting therefore samples the sector without selectively emphasizing coherent sign regions of the operator profile.

\begin{figure}[t]
\centering
\includegraphics[width=\columnwidth]{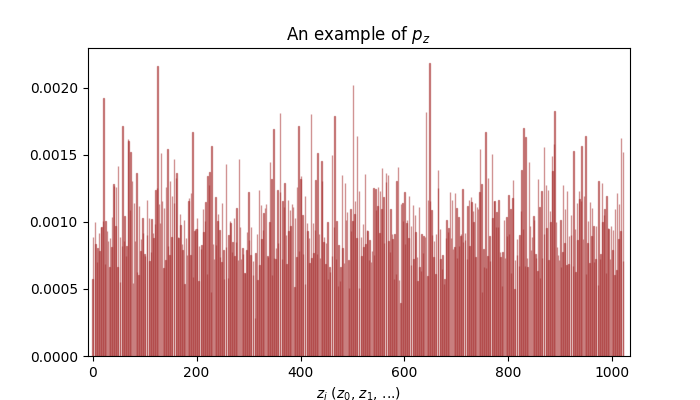}
\caption{\label{fig:haar_pz}
Example of the basis-resolved weights $p_z=|\langle r|P_{\uparrow}|z\rangle|^2$ for a Haar-random initial state.
The distribution exhibits no localized enhancement across the computational basis, illustrating the structureless weighting with no operator-aligned localization, characteristic of Haar-like sampling.}
\end{figure}

\subsubsection{Operator profiles: sign patterns from parity rules}

For diagonal Pauli-$Z$ strings, the operator profile $a_z(0)=\langle z|A(0)|z\rangle$ follows parity rules fixed by the operator support. These rules produce structured sign-alternating patterns across the computational-basis ordering.

\begin{figure*}[t]
\centering
\includegraphics[width=0.95\textwidth]{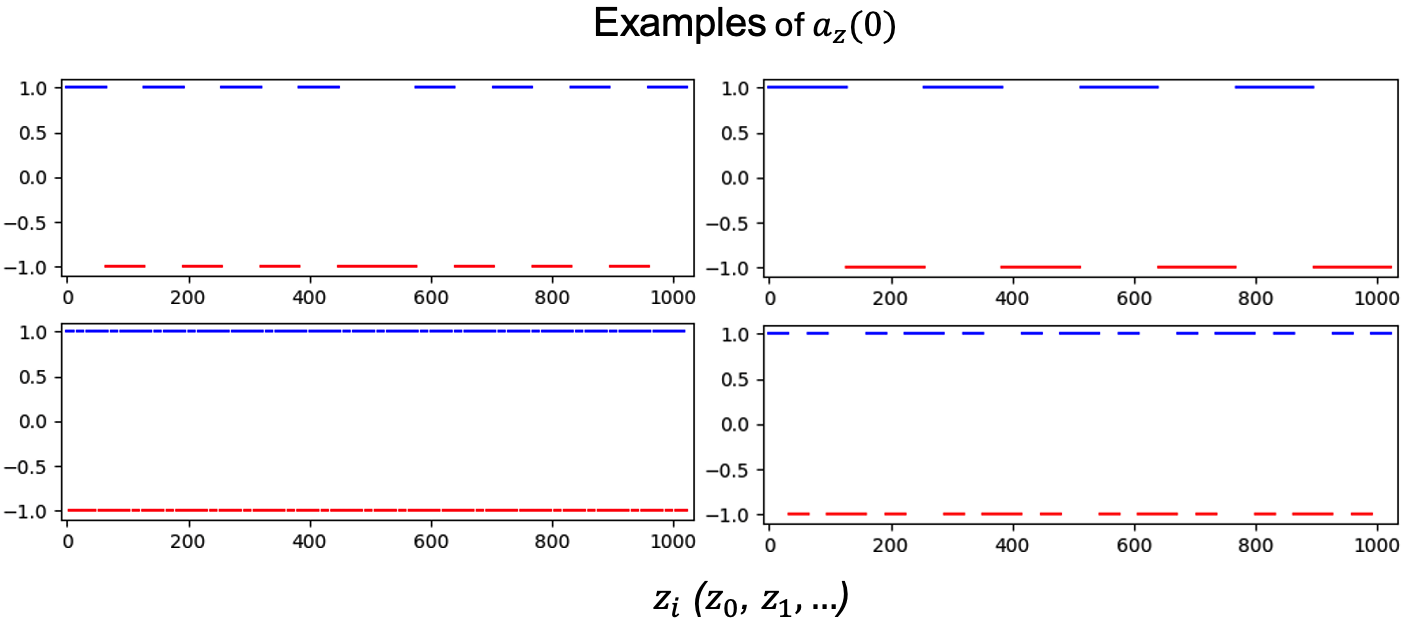}
\caption{\label{fig:az_examples}
Examples of diagonal operator profiles $a_z(0)=\langle z|A(0)|z\rangle$ for representative Pauli-$Z$--type observables.
While these operators have well-defined support, their basis-resolved profiles exhibit structured, sign-alternating patterns across $z$. For multi-qubit operators, this structure becomes increasingly rich, reflecting parity and support-dependent correlations over the computational basis. Under near-uniform weighting, however, these structured positive and negative contributions are averaged on equal footing, leading to extensive cancellation.}
\end{figure*}

\subsubsection{Weighted contributions and cumulative cancellation}

\begin{figure}[t]
\centering
\includegraphics[width=\columnwidth]{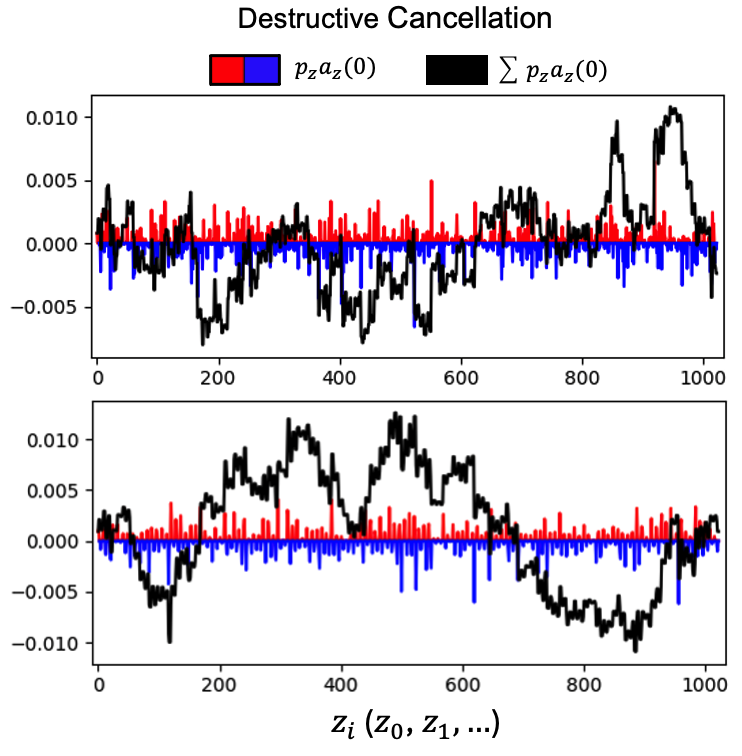}
\caption{\label{fig:haar_cancel}
Visualization of destructive cancellation under Haar-like weighting.
Bars show individual signed contributions $p_z a_z(0)$, while the black curve shows the cumulative sum $S_\uparrow(i)=\sum_{j\le i}p_{z_j}a_{z_j}(0)$.
For the illustrated sampled state, the positive and negative contributions largely cancel. The endpoint $W_\uparrow$ is a state-dependent quantity; the corresponding Haar mean is set by the projected trace.}
\end{figure}

For the sign-balanced projected profile illustrated in Fig.~\ref{fig:haar_cancel}, the mean-uniform weights produce a cumulative sum $S_\uparrow(i)$ with little net accumulation. The Haar mean of its endpoint is $\mathrm{Tr}[P_{\uparrow}A(0)]/D$, with $D=2^n$, and the sampled-state residual fluctuates around that value. The full correlator also includes the second term in Eq.~\eqref{eq:Ce_shorttime}. The appearance of a cumulative curve depends on the basis ordering; its small endpoint in this example characterizes cancellation in the prepared state, while Sec.~\ref{subsec:trace_sampling} distinguishes this from finite-shot estimation error.

\subsubsection{Interpreting the $\langle r|A(0)|r\rangle$ term in the Haar baseline}

The second term in Eq.~\eqref{eq:Ce_shorttime} originates from the $-I$ component in the decomposition $Z=2P_{\uparrow}-I$ and is required to preserve the intended ``$\pm$-sector contrast'' of a Pauli-$Z$ correlator.
For Haar-random states and traceless diagonal observables, $\mathbb{E}[\langle r|A(0)|r\rangle]=0$ exactly. Individual exact expectations fluctuate around this mean with the Haar variance described in Sec.~\ref{subsec:trace_sampling}. For engineered ensembles, the mean can change, so the average term and the projected contribution must be retained together.

\subsection{\label{sec:peaked_breakdown}Peaked ensembles: localized weighting and survival of structure}

In the Haar baseline, the mean ensemble weights are uniform and do not favor an operator-sign sector.
Peaked ensembles replace this reference weighting with controllable nonuniform weights, thereby changing the ensemble-dependent expectation value.
Quantum state preparation physically realizes this weighting, so each measurement shot samples a bitstring from the engineered distribution, without classical reweighting, rejection sampling, or postselection.

\subsubsection{From global averaging to localized partial sums}

Under a peaked state $|r\rangle$, probability mass is concentrated in a restricted region of the computational basis. The weighted sum is governed by the localized partial sum over the region where $p_z$ is appreciable. When this region aligns with basis components for which $a_z(0)$ has a coherent sign or systematically larger magnitude, their contributions reinforce one another and can retain operator-resolved structure suppressed by uniform averaging.

\begin{figure}[t]
\centering
\includegraphics[width=\columnwidth]{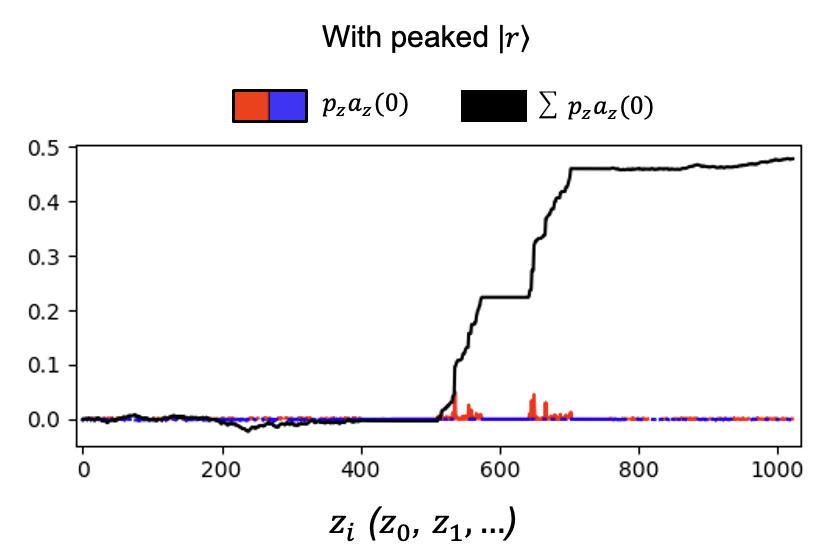}
\caption{\label{fig:peaked_cancel_break}
Example of cancellation breakdown under a peaked ensemble.
The signed contributions $p_z a_z(0)$ become concentrated in a restricted basis region, and the cumulative sum $S_\uparrow(i)$ exhibits step-like growth toward its endpoint $W_\uparrow$, indicating coherent addition of operator-aligned contributions.}
\end{figure}

\subsubsection{The average term becomes part of the contrast in engineered ensembles}

In peaked ensembles, $\langle r|A(0)|r\rangle$ changes with the same basis weights as the projected contribution.
To expose the shared structure, define the full computational-basis weights
\begin{equation}
q_z \equiv |\langle z|r\rangle|^2,\qquad \sum_z q_z = 1.
\end{equation}
If $P_{\uparrow}$ acts on a fixed qubit $k$, then
\begin{equation}
p_z=\delta_{z_k,0}\,q_z.
\end{equation}
Introducing the complementary projector $P_{\downarrow}=I-P_{\uparrow}$ gives
\begin{equation}
p^{(\downarrow)}_z \equiv \big|\langle r|P_{\downarrow}|z\rangle\big|^2
=\delta_{z_k,1}\,q_z,
\qquad
p_z + p^{(\downarrow)}_z = q_z.
\end{equation}
With these definitions, the ensemble-dependent correlator admits the equivalent two-sector contrast form
\begin{equation}
\label{eq:Ce_twosector}
C_{\mathcal{E}}(0) = \sum_{z_k=0} q_z a_z(0) - \sum_{z_k=1} q_z a_z(0),
\end{equation}
which makes explicit that the signal is governed by the difference between the $z_k=0$ and $z_k=1$ sectors.
For diagonal $A(0)$, Eq.~\eqref{eq:Ce_twosector} is equivalently
\begin{equation}
C_{\mathcal{E}}(0) = \langle r|A(0) Z_k |r\rangle.
\end{equation}
Peaked-state engineering reshapes the full basis weight distribution $q_z$ and thereby controls both terms in Eq.~\eqref{eq:Ce_shorttime}. To enhance the sector contrast in Eq.~\eqref{eq:Ce_twosector}, it concentrates weight on coherently contributing regions in the desired sector and suppresses competing contributions in the opposite sector.

\subsection{\label{sec:peaked_metrics}Sector-resolved diagnostics of peakedness and contrast}

We focus on diagnostics that enter the full ensemble-dependent correlator. Because Eq.~\eqref{eq:Ce_shorttime} contains both a projected contribution and an ensemble-dependent average, we track the masses and signed sums of both sectors. These quantities describe how $\langle r|A(0)|r\rangle$ and $C_{\mathcal E}(0)$ are jointly shaped by preparation. Finer within-sector localization diagnostics will be introduced when comparing explicit circuit families.

\paragraph{Projected-sector mass (scale).}
Since the projected weights $\{p_z\}$ are not normalized, we define their total mass
\begin{align}
\pi_{\uparrow}
&\equiv \sum_z p_z
= \langle r|\Pi_{\uparrow}|r\rangle, \notag\\
\pi_{\downarrow}
&\equiv \sum_z p^{(\downarrow)}_z
= \langle r|\Pi_{\downarrow}|r\rangle
= 1-\pi_{\uparrow}.
\label{eq:sector_masses}
\end{align}
Here $\Pi_{\uparrow,\downarrow}$ denote the projector operators onto the
$z_k=0,1$ sectors, while $\pi_{\uparrow,\downarrow}$ are the corresponding scalar sector masses.
These quantities quantify how much probability weight the prepared state places in each $Z_k$ sector, and they set the overall scale of the sector-resolved contributions defined below.

\paragraph{Sector-resolved signed sums and contrast.}
Define the sector-resolved signed sums
\begin{align}
W_{\uparrow}
&\equiv \sum_{z_k=0} q_z\,a_z(0)
= \sum_z p_z\,a_z(0), \notag\\
W_{\downarrow}
&\equiv \sum_{z_k=1} q_z\,a_z(0)
= \sum_z p^{(\downarrow)}_z\,a_z(0).
\label{eq:sector_signed_sums}
\end{align}
These quantities capture the net signed contribution accumulated within each sector under the engineered ensemble. It is useful to separate the total probability mass in each sector from the degree of intra-sector alignment with the operator profile. For nonzero sector mass, define
\begin{equation}
m_{\uparrow}
\equiv
\frac{1}{\pi_{\uparrow}}
\sum_{z_k=0} q_z\,a_z(0),
\qquad
m_{\downarrow}
\equiv
\frac{1}{\pi_{\downarrow}}
\sum_{z_k=1} q_z\,a_z(0).
\label{eq:sector_means}
\end{equation}
Equivalently,
\begin{equation}
W_{\uparrow}=\pi_{\uparrow}m_{\uparrow},
\qquad
W_{\downarrow}=\pi_{\downarrow}m_{\downarrow}.
\label{eq:sector_mass_alignment}
\end{equation}
Here $\pi_{\uparrow,\downarrow}$ quantify where probability mass is placed, while $m_{\uparrow,\downarrow}$ quantify how coherently the diagonal operator profile adds within each sector. Thus, sector polarization alone does not determine the contrast.

The two key combinations appearing in the full estimator are then
\begin{equation}
\langle r|A(0)|r\rangle
=
\sum_z q_z a_z(0)
=
W_{\uparrow}+W_{\downarrow},
\label{eq:Aavg_sector}
\end{equation}
while Eq.~\eqref{eq:Ce_twosector} gives
\begin{equation}
C_{\mathcal E}(0)
=
W_{\uparrow}-W_{\downarrow}
=
\pi_{\uparrow}m_{\uparrow}
-
\pi_{\downarrow}m_{\downarrow}.
\label{eq:CE_sector}
\end{equation}
Thus $\langle r|A(0)|r\rangle$ is the ensemble-dependent \emph{sum} of the same sector contributions that define the \emph{contrast}. Equations~\eqref{eq:Aavg_sector} and \eqref{eq:CE_sector} therefore make explicit how state-preparation choices simultaneously shape both terms in Eq.~\eqref{eq:Ce_shorttime}. In particular, for diagonal observables with $|a_z(0)|\approx 1$, which is common for Pauli-$Z$--type operators, $\pi_{\uparrow,\downarrow}$ determine the sector populations, $m_{\uparrow,\downarrow}$ determine the intra-sector signed alignment, and $W_{\uparrow,\downarrow}$ combine both effects into the sector-resolved signed contributions.

\paragraph{Cumulative sums and their endpoints.}
For the same integer ordering as in Eq.~\eqref{eq:cumsum_def}, define the full-distribution and correlator cumulative sums by
\[
\begin{aligned}
S_A(i)&\equiv\sum_{j=0}^{i}q_{z_j}a_{z_j}(0),\\
S_C(i)&\equiv\sum_{j=0}^{i}q_{z_j}a_{z_j}(0)(-1)^{(z_j)_k}\\
&=2S_\uparrow(i)-S_A(i),
\end{aligned}
\]
where $(z_j)_k$ is bit $k$ of the basis string $z_j$. For diagonal $A(0)$ and $B(0)=Z_k$, the endpoints at $D=2^n$ are
\[
\begin{aligned}
S_\uparrow(D-1)&=W_\uparrow,\\
S_A(D-1)&=W_\uparrow+W_\downarrow=\langle r|A(0)|r\rangle,\\
S_C(D-1)&=W_\uparrow-W_\downarrow=C_{\mathcal E}(0).
\end{aligned}
\]
Thus each cumulative curve identifies where its specified signed contributions accumulate, with the full correlator determined by the sector difference. The same sum-and-difference identities hold for empirical weights estimated from measurement outcomes.

\paragraph{Relation to earlier projected-sector diagnostics.}
The support-overlap (denoted $S_A$ in that work) and biased-expectation ratio $E_A$ in Sec.~III D of Ref.~\cite{Kim2025ADD} already distinguish projected support from signed alignment. Using the same projected weights, with $|a_z(0)|=1$ and $\pi_{\uparrow}>0$, their support-overlap equals $\pi_{\uparrow}$ and $E_A=W_{\uparrow}/\pi_{\uparrow}=m_{\uparrow}$; the projected signal used there corresponds to $W_{\uparrow}$. The earlier work also displayed the projected-plus-average expression before focusing on the projected contribution. Here we retain both contributions throughout the correlator analysis: the complementary-sector quantities complete the sum in Eq.~\eqref{eq:Aavg_sector} and the difference in Eq.~\eqref{eq:CE_sector}.

\paragraph{Information supplied by the sector diagnostics.}
At fixed sector mass, different conditional alignments can give different contrasts. Even fixing both selected-sector diagnostics leaves the full correlator undetermined. For $A(0)=Z_j$, $B(0)=Z_k$, and $j\ne k$, consider two distributions in the basis order $(z_k,z_j)=(00,01,10,11)$:
\begin{equation}
\begin{aligned}
q^{(1)}&=(0.45,\,0.15,\,0.30,\,0.10),\\
q^{(2)}&=(0.45,\,0.15,\,0.10,\,0.30).
\end{aligned}
\label{eq:selected_sector_example}
\end{equation}
Both have $\pi_{\uparrow}=0.6$, $m_{\uparrow}=0.5$, and $W_{\uparrow}=0.3$. Their complementary-sector mass is also identical, $\pi_{\downarrow}=0.4$, but $m_{\downarrow}=0.5$ and $-0.5$, respectively. Thus $W_{\downarrow}=0.2$ and $-0.2$, giving $C_{\mathcal E}(0)=0.1$ and $0.5$, while $\langle r|A(0)|r\rangle=0.5$ and $0.1$. The selected-sector metrics agree, yet the full correlator and the ensemble average differ. Reporting both sectors identifies the information missing from the projected signal alone.

\paragraph{Accuracy of a selected-sector approximation.}
For normalized $q_z$ and a real diagonal profile satisfying $|a_z(0)|\leq1$, the triangle inequality gives
\begin{equation}
\begin{aligned}
|C_{\mathcal E}(0)-W_{\uparrow}|
&=|W_{\downarrow}|\\
&\leq \sum_{z_k=1}q_z|a_z(0)|\leq\pi_{\downarrow}.
\end{aligned}
\label{eq:omitted_sector_bound}
\end{equation}
Consequently, $W_{\uparrow}-\pi_{\downarrow}\leq C_{\mathcal E}(0)\leq W_{\uparrow}+\pi_{\downarrow}$. Near-unit selected-sector occupation guarantees a small absolute correction from the omitted sector. When its mass is appreciable, measuring $W_{\downarrow}$ can give a much tighter characterization, since positive and negative contributions may cancel within that sector. This elementary bound quantifies the deterministic omission of a sector; finite-shot uncertainty is a separate issue. The same algebra holds for an empirical distribution, with all quantities replaced by their empirical counterparts.

\subsection{\label{sec:sec3_summary}Summary and implications for circuit design}

The Haar baseline illustrates how operator contributions combine under mean-uniform weighting, including cancellations consistent with the infinite-temperature reference.
Peaked ensembles change this weighting and the corresponding correlator by concentrating probability on selected basis contributions.
In engineered ensembles, the ``average'' term $\langle r|A(0)|r\rangle$ is no longer generically negligible and should be interpreted together with the projected contribution as part of a two-sector contrast mechanism.
The quantities $\pi_{\uparrow,\downarrow}$, $m_{\uparrow,\downarrow}$, and $W_{\uparrow,\downarrow}$ therefore provide a practical sector-resolved language for quantifying how state preparation redistributes probability mass, controls intra-sector alignment, and shapes both the contrast and the ensemble-dependent average. Finer concentration properties will be characterized below using explicit weight-distribution diagnostics for the circuit families under study.
In the next section, we turn to explicit circuit constructions that generate such peaked ensembles and allow one to tune this contrast in practice under realistic depth and noise constraints.

\section{Circuit constructions for peaked ensembles}
\label{sec:peaked_circuits}

The Grover-based and shallow circuit families follow the peaked-state preparation strategies introduced in Ref.~\cite{Kim2025ADD}. Here they provide controlled preparations for the two-sector analysis of Secs.~II--III. With the operator profile specified, the circuit controls reshape the computational-basis weights and hence the masses and signed contributions that determine the prepared-ensemble correlator. We use these constructions to connect preparation choices to the sector diagnostics.

\subsection{Design objective and interface}
\label{subsec:design_objective}

\paragraph{Weights, operator profiles, and sector contrast.}

The circuit design objective follows directly from the basis and
sector-resolved diagnostics introduced in Sec.~\ref{sec:sec3}. We use the basis weights $q_z$, diagonal operator profile $a_z(0)$, sector masses
$\pi_{\uparrow,\downarrow}$, intra-sector signed means
$m_{\uparrow,\downarrow}$, and sector-resolved signed sums
$W_{\uparrow,\downarrow}$ defined there. For notational simplicity,
throughout this section we write $a_z\equiv a_z(0)$.

With this notation, the ensemble-dependent diagonal correlator used as
the design target is the sector imbalance
\begin{equation}
C_{\mathcal E}(0)
=
W_{\uparrow}-W_{\downarrow}
=
\pi_{\uparrow}m_{\uparrow}
-
\pi_{\downarrow}m_{\downarrow}.
\label{eq:contrast}
\end{equation}
The design task is to shape the prepared basis weights by jointly controlling probability mass and intra-sector alignment to increase the sector contrast.

\paragraph{Operational measurement pipeline (diagonal case).}
Operationally, each shot returns a bitstring $z$.
For diagonal observables, $a_z$ is computed classically and shots are binned by the sector label $z_k$.
With $N_{\mathrm{sh}}$ shots, one may form
\begin{equation}
\begin{aligned}
\widehat{W}_\uparrow
&= \frac{1}{N_{\mathrm{sh}}}
\sum_{s=1}^{N_{\mathrm{sh}}}
a_{z^{(s)}}\,\mathbf{1}[z^{(s)}_k=0], \\
\widehat{W}_\downarrow
&= \frac{1}{N_{\mathrm{sh}}}
\sum_{s=1}^{N_{\mathrm{sh}}}
a_{z^{(s)}}\,\mathbf{1}[z^{(s)}_k=1].
\end{aligned}
\label{eq:W_estimators}
\end{equation}
so that the shot-based estimator of the ensemble-dependent correlator is
\begin{equation}
\widehat{C}_{\mathcal E}(0) = \widehat{W}_\uparrow-\widehat{W}_\downarrow.
\label{eq:CE_hat_estimator}
\end{equation}
This is the operational quantity used in both the noiseless numerical simulations of Sec.~V and the IBM-hardware demonstrations of Sec.~VI.

\paragraph{Input/output interface.}
The inputs are an explicit description of the observable and its relevant bit or parity rules, a desired sector/sign bias, and hardware constraints on depth, two-qubit gates, and connectivity. The operator qubit support $\mathcal S_A=\mathrm{supp}(A)$ specifies which qubits it acts on. This differs from the basis states where $a_z\ne0$ and from the target set $\mathcal G$ selected for preparation. For a Pauli-$Z$ string, $a_z=\pm1$ on every basis state; its parity structure supplies the information used to choose $\mathcal G$. The output is a circuit family with specified parameters whose sector contributions are measured. The examples below make the required structural information and preparation costs explicit.

\paragraph{Two complementary roles.}
We consider two circuit families.
The first is an \emph{oracle-based Grover-type} construction, which serves as a structure-aligned benchmark with a clear idealized upper bound but can be depth- and noise-sensitive.
The second is an \emph{oracle-free shallow} construction, which may have more limited idealized peakedness at fixed depth but is more portable and typically more robust in larger-qubit NISQ settings.

\subsection{Oracle-based structured peaking: Grover-type construction}
\label{subsec:grover}

We use Grover-type amplitude amplification \cite{grover1997needle} to transfer probability mass into a designated good subspace. This provides a \emph{structure-aligned benchmark} for controlling ensemble weights and their signed contributions.

\subsubsection{Good-set definition without exhaustive evaluation of $a_z$}
\label{subsubsec:grover_goodset}

For the structured observables considered here, a target can be specified from a compact rule. Consider $A(0)=\bigotimes_{j\in\mathcal S_A}Z_j$, with a known nonempty support of size $s=|\mathcal S_A|$, and $B(0)=Z_k$ with $k\notin\mathcal S_A$. For a chosen parity $b\in\{0,1\}$, define
\begin{equation}
\begin{aligned}
a_z&=(-1)^{\sum_{j\in\mathcal S_A}z_j},\\
\mathcal G_b&=\{z:z_k=0,\ \bigoplus_{j\in\mathcal S_A}z_j=b\}.
\end{aligned}
\label{eq:structured_target}
\end{equation}
Every member of $\mathcal G_b$ contributes the same sign $(-1)^b$ to the diagonal product $A(0)B(0)$. Reading the support list and specifying this predicate takes $O(s)$ indexed operations; testing membership uses $O(s)$ bit operations. Neither step enumerates the $2^n$ basis states. The independent sector and parity constraints give $|\mathcal G_b|/2^n=1/4$. This is an analytic design example; the reported benchmark settings are specified separately. Fixed-bit patterns and explicit conjunctions of a small number of such rules give further structured targets.

A key design parameter is the good-set fraction
\begin{align}
f \equiv \frac{|G|}{2^n}.
\end{align}
If $f$ is too large, often close to $1/2$ for simple parity-based splits, concentration into a small number of states is structurally limited.
If $f$ is too small, the required number of Grover iterations becomes too costly for NISQ depth budgets.
We therefore work in the regime of small feasible iteration counts $T$ and design $f$ accordingly.
As an idealized guide, the standard two-dimensional Grover picture suggests
\begin{align}
f_{\mathrm{target}}(T)\approx \sin^2\!\Big(\frac{\pi}{4T+2}\Big),
\label{eq:f_target}
\end{align}
under uniform initial mixing and matched ideal reflections, by placing the first amplification maximum at iteration $T$~\cite{brassard2002amplitude}. We use Eq.~\eqref{eq:f_target} as a design guide; integer iteration counts and hardware constraints determine the feasible operating point. In the small-$f$ regime, the usual constant-success amplification scale is $T=O(f^{-1/2})$. Thus a cheap membership predicate can still entail many circuit repetitions within a preparation when the target fraction is small.

\subsubsection{Circuit template and implementation notes}
\label{subsubsec:grover_template}

After an initial mixing layer, the Grover-type circuit applies $O_f$ followed by diffusion $D$ for each of $T$ iterations. For the predicate in Eq.~\eqref{eq:structured_target}, accumulate the $s$-bit parity on one clean ancilla, apply a controlled phase for the required parity and $z_k=0$, and uncompute. This uses $2s$ CNOT gates, one CZ, and a constant number of $X$ gates, returning the ancilla to zero. These are logical counts before routing. Fixed-bit conjunctions instead require a multi-controlled phase and its chosen decomposition.

The diffusion is a reflection about the initial superposition. Counting logical gates consistently for each block gives
\begin{equation}
g_{\mathrm{prep}}=g_{\mathrm{init}}
+T\bigl(g_{\mathrm{oracle}}+g_{\mathrm{diff}}\bigr).
\label{eq:grover_resource_accounting}
\end{equation}
Here $g_{\mathrm{diff}}$ includes the synthesis of the reflection, whose cost depends on available ancillas and gate decompositions. The oracle cost alone does not give the preparation cost. Routing, compiled depth, and repetitions for measurement must also be accounted for; the analytic predicate example does not specify the transpiled resources of the hardware benchmarks.

\begin{algorithm}[t]
\caption{Grover-type peaking (oracle-based)}
\label{alg:grover_peaking}
\begin{algorithmic}[1]
\Require $n$-qubit register; explicit observable description and bit/parity rule; desired sign; iterations $T$; synthesis resources
\Ensure prepared state $|r\rangle$ for sector-resolved evaluation
\State Specify $\mathcal G=\{z:\chi_{\mathcal G}(z)=1\}$ from the supplied rule, as in Eq.~\eqref{eq:structured_target}; account separately for any additional target search

\State Construct the phase oracle $O_{\mathcal G}: |z\rangle \mapsto (-1)^{\chi_{\mathcal G}(z)}|z\rangle$
\State Initialize $|0\rangle^{\otimes n}$
\State Apply an initial mixing layer (e.g., $H^{\otimes n}$)
\For{$t=1,\dots,T$}
  \State Apply $O_{\mathcal G}$
  \State Apply diffusion $D$
\EndFor
\State \Return $|r\rangle$
\end{algorithmic}
\end{algorithm}

\begin{figure}[t]
  \centering
  \scalebox{0.85}{
  \Qcircuit @C=1.0em @R=0.2em @!R { \\
    \lstick{q_0} & \gate{H} \barrier[0em]{9} & \qw & \multigate{9}{O_f} & \multigate{9}{D} \barrier[0em]{9} & \qw & \multigate{9}{O_f} & \multigate{9}{D} \barrier[0em]{9} & \qw & \multigate{9}{O_f} & \multigate{9}{D} & \qw & \qw \\
    \lstick{q_1} & \gate{H}                  & \qw & \ghost{O_f}       & \ghost{D}                  & \qw & \ghost{O_f}       & \ghost{D}                  & \qw & \ghost{O_f}       & \ghost{D}       & \qw & \qw \\
    \lstick{q_2} & \gate{H}                  & \qw & \ghost{O_f}       & \ghost{D}                  & \qw & \ghost{O_f}       & \ghost{D}                  & \qw & \ghost{O_f}       & \ghost{D}       & \qw & \qw \\
    \lstick{q_3} & \gate{H}                  & \qw & \ghost{O_f}       & \ghost{D}                  & \qw & \ghost{O_f}       & \ghost{D}                  & \qw & \ghost{O_f}       & \ghost{D}       & \qw & \qw \\
    \lstick{q_4} & \gate{H}                  & \qw & \ghost{O_f}       & \ghost{D}                  & \qw & \ghost{O_f}       & \ghost{D}                  & \qw & \ghost{O_f}       & \ghost{D}       & \qw & \qw \\
    \lstick{q_5} & \gate{H}                  & \qw & \ghost{O_f}       & \ghost{D}                  & \qw & \ghost{O_f}       & \ghost{D}                  & \qw & \ghost{O_f}       & \ghost{D}       & \qw & \qw \\
    \lstick{q_6} & \gate{H}                  & \qw & \ghost{O_f}       & \ghost{D}                  & \qw & \ghost{O_f}       & \ghost{D}                  & \qw & \ghost{O_f}       & \ghost{D}       & \qw & \qw \\
    \lstick{q_7} & \gate{H}                  & \qw & \ghost{O_f}       & \ghost{D}                  & \qw & \ghost{O_f}       & \ghost{D}                  & \qw & \ghost{O_f}       & \ghost{D}       & \qw & \qw \\
    \lstick{q_8} & \gate{H}                  & \qw & \ghost{O_f}       & \ghost{D}                  & \qw & \ghost{O_f}       & \ghost{D}                  & \qw & \ghost{O_f}       & \ghost{D}       & \qw & \qw \\
    \lstick{q_9} & \gate{H}                  & \qw & \ghost{O_f}       & \ghost{D}                  & \qw & \ghost{O_f}       & \ghost{D}                  & \qw & \ghost{O_f}       & \ghost{D}       & \qw & \qw \\
  }}
  \caption{Grover-type peaking circuit with boxed oracle and diffusion operators.}
  \label{fig:grover}
\end{figure}

\paragraph{Weight visualization.}
The induced-weight plots complement the signed-contribution diagnostics in Sec.~\ref{sec:sec3}.
Let $q_{(1)}\ge q_{(2)}\ge\cdots$ denote the weights $q_z$ sorted in descending order.
The cumulative mass
\begin{align}
M(K)=\sum_{j=1}^{K} q_{(j)}
\label{eq:cum_mass}
\end{align}
quantifies how much total probability is concentrated in the top $K$ basis states.

\begin{figure}[t]
  \centering
  \includegraphics[width=\linewidth]{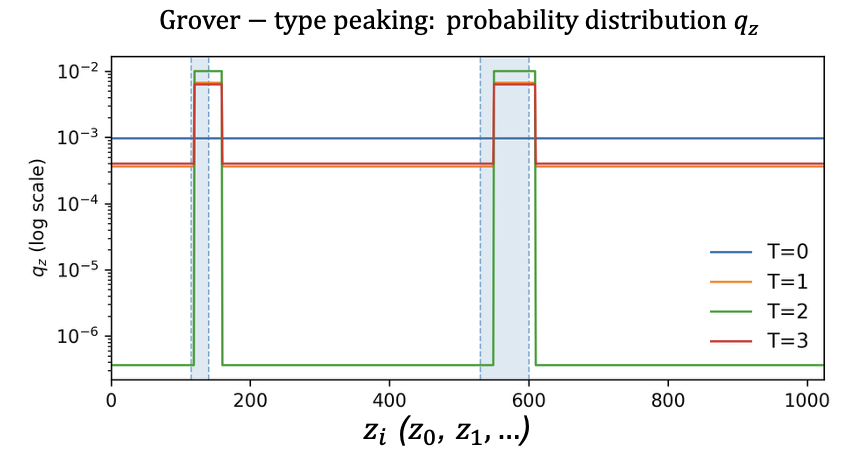}
  \caption{Probability weights $q_z = |\langle z|r\rangle|^2$ as a function of the
  computational-basis index $z$ for different Grover iteration counts $T$ (simulation).
  Shaded regions indicate the marked intervals $G$.}
  \label{fig:peaked_grover}
\end{figure}

\subsubsection{Weight distributions produced by Grover-type peaking}
\label{subsubsec:grover_weights}

Figure~\ref{fig:peaked_grover} shows the computational-basis weights $q_z$ produced by Grover-type peaking for several iteration counts $T$.
At $T=0$, the initial mixing layer yields an approximately uniform distribution, $q_z\simeq 2^{-n}$.
As $T$ increases, amplitude amplification transfers probability mass into $G$ and depletes the unmarked complement.
The resulting weight profile develops a characteristic plateau structure, with distinct levels inside and outside the marked subspace.
Because the dynamic range of $q_z$ grows rapidly with $T$, this structure is most clearly revealed on a logarithmic scale.

In the ideal two-dimensional Grover picture, the distribution remains approximately uniform \emph{within} each subspace, leading to
\begin{align}
q_z \approx
\begin{cases}
P_G(T)/|G|, & z\in G,\\[2pt]
\bigl(1-P_G(T)\bigr)/(2^n-|G|), & z\notin G,
\end{cases}
\end{align}
with $P_G(T)=\sin^2\!\bigl((2T+1)\theta\bigr)$ and $\sin^2\theta=f$.
Accordingly, the separation between the inside-$G$ and outside-$G$ plateaus, together with its dependence on $T$, provides a direct visualization of peakedness.

The union of the marked intervals forms a single good subspace $G$ that undergoes a global Grover rotation, including when those intervals are disjoint in the chosen basis ordering. In the ideal symmetric model, the weight is approximately uniform over all $z\in G$, so equal-height marked plateaus are expected. Interval-to-interval non-uniformity should therefore be interpreted as arising from non-ideal oracle/diffusion synthesis, transpilation, or noise.

Finally, the plateau heights need not vary monotonically with $T$, reflecting the oscillatory nature of amplitude amplification.
In practice, we therefore treat $T$ as a controllable knob and explore only a small, practically accessible range of iterations.

\subsubsection{Why Grover-type peaking can fail in practice}
\label{subsubsec:grover_smalln}

Grover-type peaking may appear weaker than expected, or may fail to produce a useful contrast gain, for several reasons:
\begin{enumerate}
  \item \textbf{The good-set fraction $f$ may be too large.}
  Simple sign/parity predicates often yield $f\approx 1/2$, which limits concentration into a small number of states.
  \item \textbf{Increasing $P(G)$ need not increase contrast.}
  Our target is the sector contrast $C_{\mathcal E}(0)=W_\uparrow-W_\downarrow$, not $P(G)$ itself.
  If the predicate is weakly correlated with the sector label $z_k$, Grover amplification may increase the total weight on $G$ without generating a comparable change in contrast.
  \item \textbf{$T$ is both discrete and noise-sensitive.}
  The ideal good-set mass oscillates as
  \begin{align}
    P_G(T)=\sin^2\big((2T+1)\theta\big),
  \end{align}
  so feasible iteration counts provide only a coarse control knob.
  At the same time, multi-controlled oracle and diffusion synthesis introduces depth, routing overhead, and coherent error that readily distort the intended rotation.
\end{enumerate}
For these reasons, we use Grover-type peaking primarily as a structure-revealing benchmark and complement it with shallow constructions designed for better NISQ robustness.

\subsection{Oracle-free shallow peaking: NISQ-friendly construction}
\label{subsec:shallow}

If Grover-type peaking provides a clean benchmark with a clear idealized control knob, oracle-free shallow peaking provides a \emph{practical low-depth} route to inducing structured non-uniform weights $q_z$ that remains viable in larger-qubit NISQ settings. The shallow construction should be viewed as implementing two distinct controls: a coarse sector-biasing step, which redistributes probability mass between projector-defined sectors, and a finer intra-sector shaping step, which redistributes probability within the selected support. The latter distinction is important: mere sector polarization can be achieved by simple product-state biasing, whereas operator-aligned cancellation control requires the induced weights to carry operator-aligned structure within the relevant sector.

\subsubsection{Construction principle: partial superposition, bias, and targeted sparse entanglement}
\label{subsubsec:shallow_principle}

A typical shallow template combines:
(i) \emph{selective Hadamards}, producing partial rather than global superposition;
(ii) \emph{single-qubit bias rotations}, such as $R_y(\theta_i)$, to tune coarse sector populations; and
(iii) \emph{targeted sparse entangling operations}, typically a small number of CZ/CNOT gates, introduced only when required to encode operator-dependent correlations (e.g., parity or multi-qubit structure).
The design objective is to induce a controlled departure from uniform averaging, producing structured, non-uniform weights aligned with the operator profile. In this view, the shallow construction implements both a coarse sector-biasing mechanism and a finer intra-sector shaping mechanism.

Let $U_{\mathrm{init}}$ denote the initial preparation, including fixed-bit settings, selective Hadamards, and any initial bias rotations. For a template with fixed shaping angles and an ordered entangling pattern, the preparation before optional randomization is
\[
|r_d\rangle=\bigl[U_{\mathcal E}R(\{\theta_i\})\bigr]^d
U_{\mathrm{init}}|0\rangle^{\otimes n},
\]
where $R(\{\theta_i\})$ applies single-qubit shaping rotations, such as $R_y(\theta_i)$, and $U_{\mathcal E}$ applies the sparse CZ/CNOT pattern. The shaping depth $d$ counts complete rotation-and-entangling layers; $d=0$ includes only $U_{\mathrm{init}}$. A depth comparison at fixed parameters holds the initial preparation, angles, and gate ordering fixed. The rotations allow the repeated layers to change the sorted computational-basis probabilities. CZ/CNOT gates alone change phases or permute basis weights and therefore preserve their sorted values.

For a qubit initially in $|0\rangle$, the rotation $R_y(\theta)=e^{-i\theta Y/2}$ gives
\[
P(0)=\cos^2(\theta/2),\qquad P(1)=\sin^2(\theta/2),
\]
which set the sector masses $\pi_\uparrow$ and $\pi_\downarrow$ at this stage when the qubit labels the two sectors. In Fig.~\ref{fig:shallow}, the active qubits enter the first shaping rotation in $|+\rangle=H|0\rangle$. Immediately after $R_y(\theta)$ and before the CNOT pattern, their probabilities are
\[
P(0)=\frac{1-\sin\theta}{2},\qquad P(1)=\frac{1+\sin\theta}{2}.
\]
Later shaping rotations act on the states produced by the preceding layers; the final computational-basis weights are determined by the full preparation circuit.

\subsubsection{Using operator structure without an oracle}
\label{subsubsec:shallow_structure}

Here \emph{oracle-free} means that preparation uses no Grover marking oracle or diffusion; classical knowledge of the operator and target rule still guides the circuit. A constructive example for $\mathcal G_b$ in Eq.~\eqref{eq:structured_target} is to keep qubit $k$ in $|0\rangle$, choose a pivot $j_0\in\mathcal S_A$ initialized to $|0\rangle$, and apply Hadamards to all other qubits. Apply CNOTs from each $j\in\mathcal S_A\setminus\{j_0\}$ to the pivot, followed by $X_{j_0}$ if $b=1$. This prepares a uniform superposition over $\mathcal G_b$ with $n-2$ Hadamards, $s-1$ CNOTs, and at most one $X$, before routing. The sequential pivot construction has depth $O(s)$ and requires no ancilla. It illustrates a tractable mapping from a known rule to a circuit, separately from the benchmark templates. The same bit/parity constraints also allow efficient classical sampling.

\subsubsection{Weight distributions produced by shallow peaking}
\label{subsubsec:shallow_weights}

\begin{figure}[t]
  \centering
  \includegraphics[width=\linewidth]{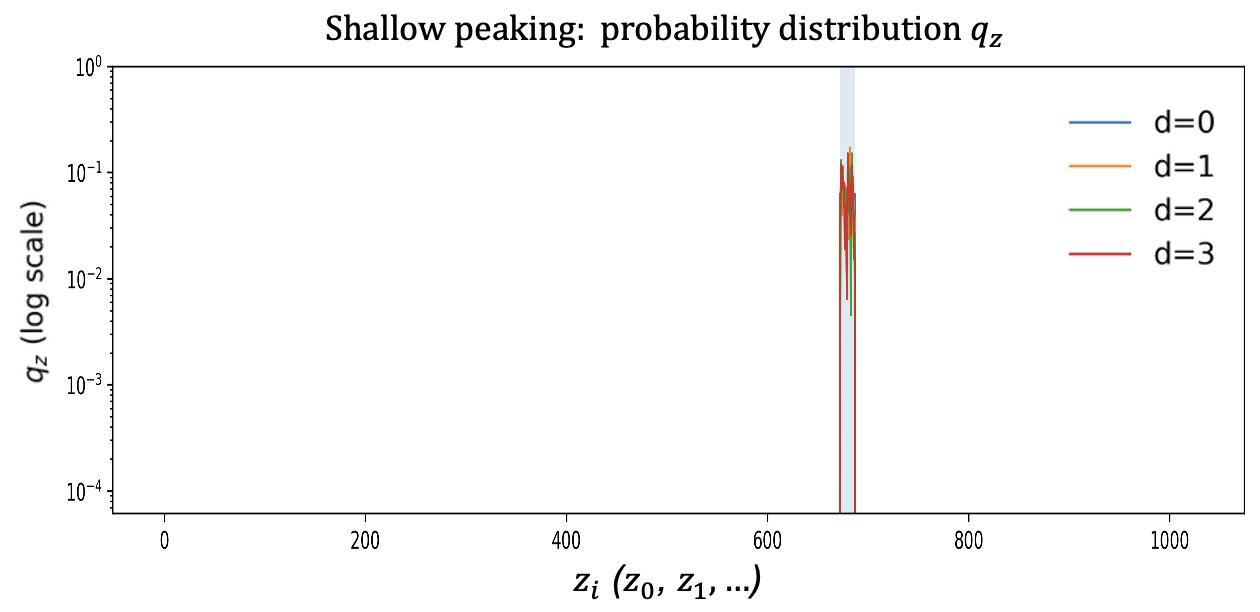}
  \caption{Probability weights $q_z = |\langle z|r\rangle|^2$ as a function of the computational-basis index $z$ for different shallow preparation depths $d$ (simulation). Shaded regions indicate the targeted block.}
  \label{fig:shallow_weights}
\end{figure}

Figure~\ref{fig:shallow_weights} shows the computational-basis weights $q_z$ induced by the oracle-free shallow construction for several shaping depths $d$.
The shaded window indicates the targeted block $G$, namely the basis region selected by the imposed bit and sector-bias pattern.

At $d=0$, the initial preparation $U_{\mathrm{init}}$ concentrates most of the probability mass within the target region $G$ for this parameter choice, producing a coarse sector-selected distribution. As $d$ increases, the repeated shaping rotations and sparse entangling gates redistribute probability within this region, generating a more structured and localized intra-sector profile. As in Fig.~\ref{fig:peaked_grover}, we use a logarithmic scale to resolve the resulting dynamic range.

Grover-type peaking ideally yields an approximately two-level distribution that is nearly uniform within the marked and unmarked subspaces. Shallow peaking allows nonuniform weights within $G$; the detailed intra-block pattern depends on the chosen angles $\{\theta_i\}$, the entangler layout, and implementation details. We treat $d$ as a practical control knob: increasing $d$ can enhance peakedness at the cost of additional single- and two-qubit gates.

\subsubsection{Knobs and tuning (pre-variational heuristics)}
\label{subsubsec:shallow_tuning}

The primary knobs are the shaping angles $\{\theta_i\}$, entangling pattern $\mathcal E$, and depth $d$, together with the initial preparation. Let $g_{\mathrm{init}}$ be the logical gate count of $U_{\mathrm{init}}$, and let each repeated layer contain $n_R$ single-qubit rotations and $|\mathcal E|$ entangling gates. The preparation then uses
\[
g_{\mathrm{prep}}=g_{\mathrm{init}}+d\bigl(n_R+|\mathcal E|\bigr)
\]
logical gates. For an initial preparation using $O(n_{\mathrm a})$ local gates on $n_{\mathrm a}$ active qubits, this becomes $O\bigl(n_{\mathrm a}+d(n_R+|\mathcal E|)\bigr)$. Optional randomization, native-gate decomposition, and hardware routing add their own costs. This count assumes the parameters and layout are supplied. Testing $Q$ settings with $N_{\mathrm{sh}}$ shots each requires $QN_{\mathrm{sh}}$ circuit executions, in addition to classical design and compilation. The restricted sweeps in Sec.~\ref{sec:simulation} characterize chosen settings; they provide no general guarantee of efficient optimization or of realizing an arbitrary target at fixed shallow depth.

\subsubsection{What it produces as an ensemble}
\label{subsubsec:shallow_ensemble}

Shallow peaking is naturally interpreted as an \emph{ensemble-generation rule}.
For example, lightweight local randomization before or after the template can increase shot-to-shot diversity while preserving the intended peakedness and alignment.
When used, such randomization is applied in a sector-preserving manner so that the intended $z_k$ bias remains intact.

\begin{algorithm}[t]
\caption{Shallow peaking (oracle-free)}
\label{alg:shallow_peaking}
\begin{algorithmic}[1]
\Require $n$-qubit register; known operator qubit support $\mathcal S_A$ and bit/parity rule; initial-preparation parameters; shaping angles $\{\theta_i\}$; entangling pattern $\mathcal E$; depth $d$; optional randomization seed
\Ensure candidate prepared state $|r\rangle$ (or ensemble) for sector-resolved evaluation
\State Specify target $\mathcal G$ from the supplied operator and sector/sign rule
\State Specify $U_{\mathrm{init}}$, shaping rotations, and an ordered feasible entangling pattern using that rule; account for classical design and parameter trials

\State Initialize $|0\rangle^{\otimes n}$
\State Apply $U_{\mathrm{init}}$: set fixed bits, apply selective Hadamards, and apply any initial bias rotations in the specified order
\For{$\ell=1,\dots,d$}
  \State Apply the single-qubit shaping rotations $R(\{\theta_i\})$
  \State Apply the sparse CZ/CNOT pattern $U_{\mathcal E}$
\EndFor
\State (Optional) Apply lightweight sector-preserving randomization that does not change the intended bias
\State Sample in the computational basis to estimate $\hat q_z$ (and, if desired, $\hat W_\uparrow,\hat W_\downarrow,\hat C_{\mathcal E}(0)$)
\end{algorithmic}
\end{algorithm}

\begin{figure}[t]
  \centering
  \resizebox{\linewidth}{!}{%
  \Qcircuit @C=1.0em @R=0.2em @!R { \\
	 	\nghost{{q}_{0} :  } & \lstick{{q}_{0} :  } & \gate{\mathrm{H}} & \gate{\mathrm{R_Y}\,(\theta_0)} & \qw & \qw & \targ & \ctrl{3} & \multigate{3}{\mathrm{ShapeLayer\,(same\,as\,left)}}_<<<{3} & \qw & \qw\\
	 	\nghost{{q}_{1} :  } & \lstick{{q}_{1} :  } & \gate{\mathrm{H}} & \gate{\mathrm{R_Y}\,(\theta_1)} & \qw & \targ & \ctrl{-1} & \qw & \ghost{\mathrm{ShapeLayer\,(same\,as\,left)}}_<<<{2} & \qw & \qw\\
	 	\nghost{{q}_{2} :  } & \lstick{{q}_{2} :  } & \gate{\mathrm{H}} & \gate{\mathrm{R_Y}\,(\theta_2)} & \targ & \ctrl{-1} & \qw & \qw & \ghost{\mathrm{ShapeLayer\,(same\,as\,left)}}_<<<{1} & \qw & \qw\\
	 	\nghost{{q}_{3} :  } & \lstick{{q}_{3} :  } & \gate{\mathrm{H}} & \gate{\mathrm{R_Y}\,(\theta_3)} & \ctrl{-1} & \qw & \qw & \targ & \ghost{\mathrm{ShapeLayer\,(same\,as\,left)}}_<<<{0} & \qw & \qw\\
	 	\nghost{{q}_{4} :  } & \lstick{{q}_{4} :  } & \qw & \qw & \qw & \qw & \qw & \qw & \qw & \qw & \qw\\
	 	\nghost{{q}_{5} :  } & \lstick{{q}_{5} :  } & \gate{\mathrm{X}} & \qw & \qw & \qw & \qw & \qw & \qw & \qw & \qw\\
	 	\nghost{{q}_{6} :  } & \lstick{{q}_{6} :  } & \qw & \qw & \qw & \qw & \qw & \qw & \qw & \qw & \qw\\
	 	\nghost{{q}_{7} :  } & \lstick{{q}_{7} :  } & \gate{\mathrm{X}} & \qw & \qw & \qw & \qw & \qw & \qw & \qw & \qw\\
	 	\nghost{{q}_{8} :  } & \lstick{{q}_{8} :  } & \qw & \qw & \qw & \qw & \qw & \qw & \qw & \qw & \qw\\
	 	\nghost{{q}_{9} :  } & \lstick{{q}_{9} :  } & \gate{\mathrm{X}} & \qw & \qw & \qw & \qw & \qw & \qw & \qw & \qw\\
\\ }}
  \caption{Oracle-free shallow preparation with $d=2$ shaping layers. The initial Hadamards and fixed-bit $X$ gates are applied once. Each shaping layer consists of the four $R_y$ rotations followed by the displayed CNOT pattern; the boxed layer repeats these rotations and CNOTs.}
  \label{fig:shallow}
\end{figure}
 
\subsection{From a circuit to an ensemble-dependent correlator}
\label{subsec:circuit_to_correlator}

Our operational measurement target is $C_{\mathcal E}(0)$ for the prepared ensemble.
Each shot consists of preparing $|r\rangle$ from a chosen circuit family and sampling a bitstring $z$ in the computational basis.
For diagonal observables, $a_z$ is computed classically and shots are binned by the sector label $z_k$ (equivalently by $\Pi_{\uparrow/\downarrow}$).
The resulting sector sums $\widehat{W}_{\uparrow,\downarrow}$ and correlator $\widehat{C}_{\mathcal E}(0)$ are computed using Eqs.~\eqref{eq:W_estimators} and \eqref{eq:CE_hat_estimator}.
The Grover-type and shallow strategies use the same sector-resolved estimators and normalization, applied to their respective prepared distributions. The expectation value $C_{\mathcal E}(0)$ depends on those distributions.

For the single-string example with support size $s$, accumulating the sector statistics takes $O(N_{\mathrm{sh}}(s+1))$ classical operations once the sampled bits are available; reading or storing full $n$-bit records adds its own data-handling cost. Only counts and signed sums are required, so this task does not reconstruct or certify the full output distribution. Full-basis probability plots and exact sorted-weight diagnostics such as $M(K)$ used in simulations are separate computations whose resources can grow with $2^n$.

\subsection{Extensions: multi-qubit diagonal case and a brief non-diagonal outlook}
\label{subsec:extensions}

\subsubsection{Multi-qubit diagonal extensions}

Ref.~\cite{Kim2025ADD} already used multi-qubit Pauli profiles as circuit-design examples. In the sector-resolved framework, $B(0)$ specifies the sector partition and $A(0)$ supplies the profile within each sector. An all-zero target on a nonempty qubit subset $S$ is selected by
\begin{align}
\Pi_{0^S}=\left(\bigotimes_{i\in S}|0\rangle\langle0|_i\right)\otimes I_{\bar S}.
\end{align}
This projector selects strings with $z_i=0$ for every $i\in S$. If used to define a binary sector observable, it gives $B(0)=2\Pi_{0^S}-I$.

Keeping $B(0)=Z_k$, one may extend $A(0)$ to Pauli-$Z$ strings,
\begin{align}
A(0)=\bigotimes_{j\in \mathrm{supp}(A)} Z_j,
\end{align}
or real linear combinations of them, with identities on unused qubits implicit. This changes $a_z(0)$ while the sector label remains $z_k$.

For a Pauli-string sector observable, choose
\[
B(0)=Z_S\equiv\prod_{i\in S}Z_i,\qquad
\Pi_\pm^{(S)}=\frac{I\pm Z_S}{2}.
\]
Its eigenvalue $b_z=(-1)^{\sum_{i\in S}z_i}$ labels the even ($+$) and odd ($-$) parity sectors. For two qubits in $S$, the even sector contains both $|00\rangle$ and $|11\rangle$, whereas $\Pi_{0^S}$ selects only $|00\rangle$ on those qubits. For diagonal $A(0)$, the sector quantities become
\[
\pi_\pm=\sum_{z:b_z=\pm1}q_z,\qquad
W_\pm=\sum_{z:b_z=\pm1}q_z a_z(0),
\]
giving $C_{\mathcal E}(0)=W_+-W_-$. The mass--alignment decomposition $W_\pm=\pi_\pm m_\pm$ applies with $m_\pm=W_\pm/\pi_\pm$ when $\pi_\pm>0$. Target design can therefore use the sector rule specified by $B(0)$ together with the sign structure of $a_z(0)$.

\paragraph{Classical preprocessing and applicability.}
For an explicitly listed sum of $L$ diagonal Pauli strings with coefficients supplied to a specified precision and supports $\mathcal S_\ell$, evaluating $a_z$ at one bitstring takes $O(\sum_{\ell=1}^{L}(|\mathcal S_\ell|+1))$ arithmetic and bit operations. Efficient pointwise evaluation can define a sign predicate, but does not by itself determine its target fraction or supply an inexpensive preparation of the corresponding distribution. The full resource assessment must include target design, reversible predicate synthesis, preparation, and parameter trials. A diagonal observable provided as a table may already require $2^n$ entries. For a complicated or time-evolved observable without a compact accessible profile, obtaining that profile or a useful target can itself be costly; the present framework does not provide a general efficient discovery algorithm. Its constructive guarantees here concern the explicitly specified bit/parity examples.

\subsubsection{Brief outlook on non-diagonal observables}
An individual Pauli string involving $X$ or $Y$ can be measured using local basis rotations, for example $U_X=H$ and $U_Y=H S^\dagger$, before computational-basis measurement. This measurement step alone does not identify a useful target for a general noncommuting observable or provide a compact description of $A(t)$. Local-estimator formulations additionally require access to the relevant matrix elements and state-amplitude ratios. Their acquisition cost and the off-diagonal contributions must be addressed in any extension beyond the present diagonal benchmarks.

\subsection{Design heuristics and trade-offs: when to use which method}
\label{subsec:tradeoff}

Grover-type peaking provides a clear idealized benchmark and a dominant control knob $T$, but is sensitive to depth, noise, and the choice of predicate, particularly the good-set fraction $f$ and its correlation with sector contrast.
Shallow peaking may have more limited idealized expressivity at fixed depth, but benefits from low depth, portability, and better practical robustness on NISQ hardware. In Secs.~\ref{sec:simulation} and~\ref{sec:hardware}, we characterize both preparation strategies using a common set of sector-resolved diagnostics for their ensemble-dependent correlators. Section~\ref{sec:simulation} examines the response to preparation controls, while Sec.~\ref{sec:hardware} evaluates representative hardware-accessible instances.

\section{Simulation Study: Parameter-Dependent Peaking and Contrast}
\label{sec:simulation}

To quantify how the preparation controls affect the diagnostics, we perform noiseless parameter sweeps for the two circuit families illustrated in Sec.~IV. We use the ensemble-dependent sector contrast defined in Eq.~\eqref{eq:contrast}, together with the sector-mass diagnostic $\pi_{\uparrow}$ and the cumulative concentration metric
\[
    M(K)=\sum_{j=1}^{K} q_{(j)},
\]
where $q_{(1)}\ge q_{(2)}\ge\cdots$ are the computational-basis weights sorted in descending order.
For the numerical sweeps, we use the empirical frequencies $\hat q_z=N_z/N_{\rm sh}$, where $N_z$ is the measured count of outcome $z$ and $N_{\rm sh}=\sum_z N_z$ is the total number of shots. Sector masses and contrasts are evaluated by substituting these frequencies into their defining sums. The corresponding empirical concentration is
\[
    \widehat M(K)=\sum_{j=1}^{K}\hat q_{(j)},
\]
with the ordering determined from the sampled frequencies. In the numerical results below and in Fig.~\ref{fig:shallow_sim_sweep}, we retain the notation $M(K)$ for this empirical statistic. Noiseless here refers to the circuit evolution; finite-shot sampling fluctuations remain. Sorting and summing the largest observed frequencies can overestimate the concentration of the underlying probability distribution.

Both sweeps use $n=10$ qubits with $A(0)=Z_7$ and $B(0)=Z_9$. For each fixed circuit, we compute its ideal computational-basis probabilities and draw 100 independent samples of $N_{\rm sh}=2048$ shots each. Circuit parameters are held fixed across repetitions, with no additional circuit randomization. Reported values are repetition means, and the quoted uncertainties and error bars are sample standard deviations across these 100 repetitions.

These zero-time diagonal benchmarks provide a controlled setting in which the operator sign pattern and uniform reference are known. The sweeps characterize how preparation changes sector occupation, signed contrast, and concentration. Because $M(K)$ depends only on sorted probabilities, it is unchanged by permutations of the basis weights and does not determine their alignment with a fixed operator profile. We therefore interpret this concentration metric alongside the sector-resolved quantities to characterize the prepared ensemble.

\subsection{Grover-type peaking: iteration-count sweep}
\label{subsec:grover_sim}

\begin{figure*}[t]
    \centering
    \begin{minipage}{0.49\textwidth}
        \centering
        \includegraphics[width=\linewidth]{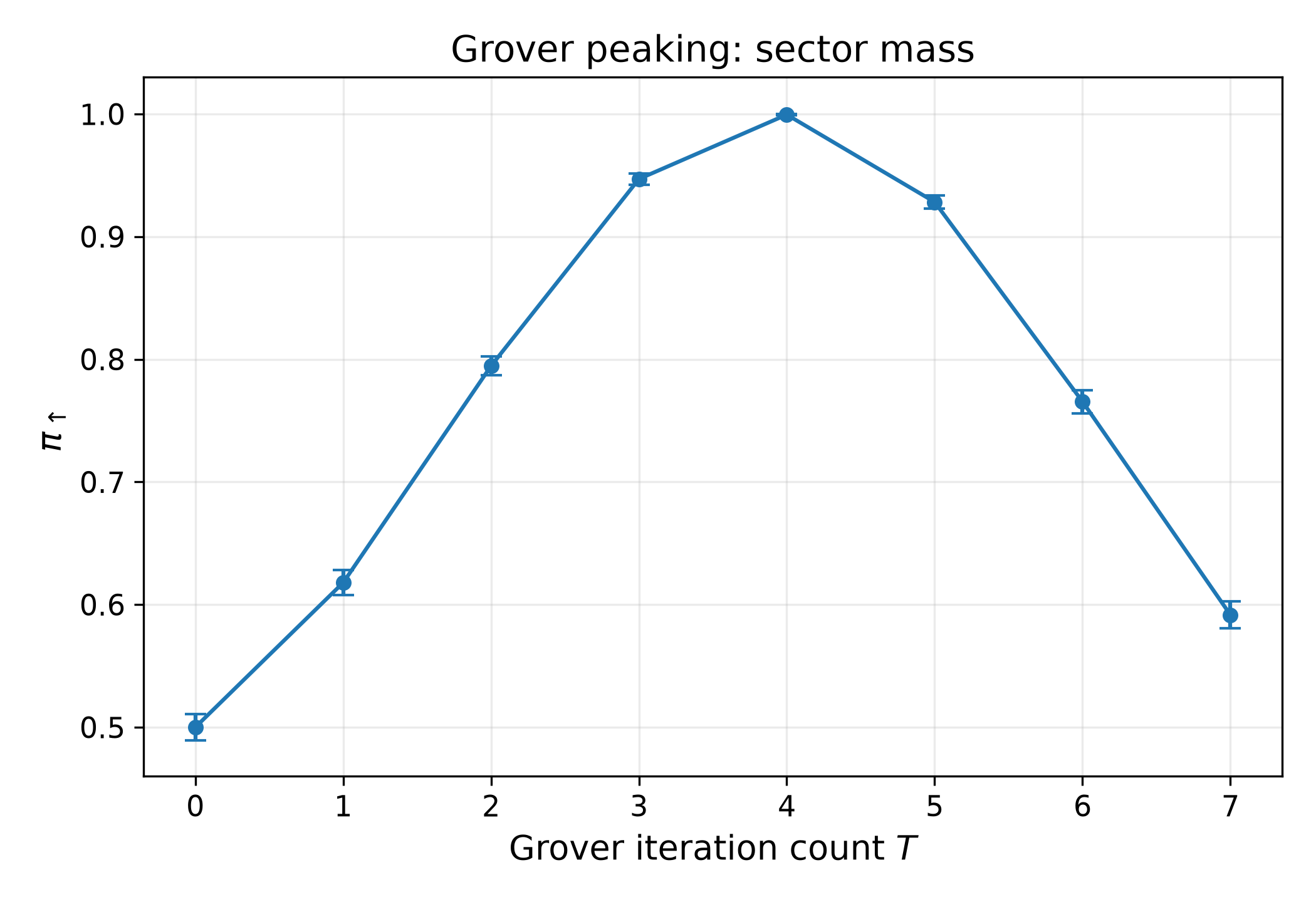}
    \end{minipage}
    \hfill
    \begin{minipage}{0.49\textwidth}
        \centering
        \includegraphics[width=\linewidth]{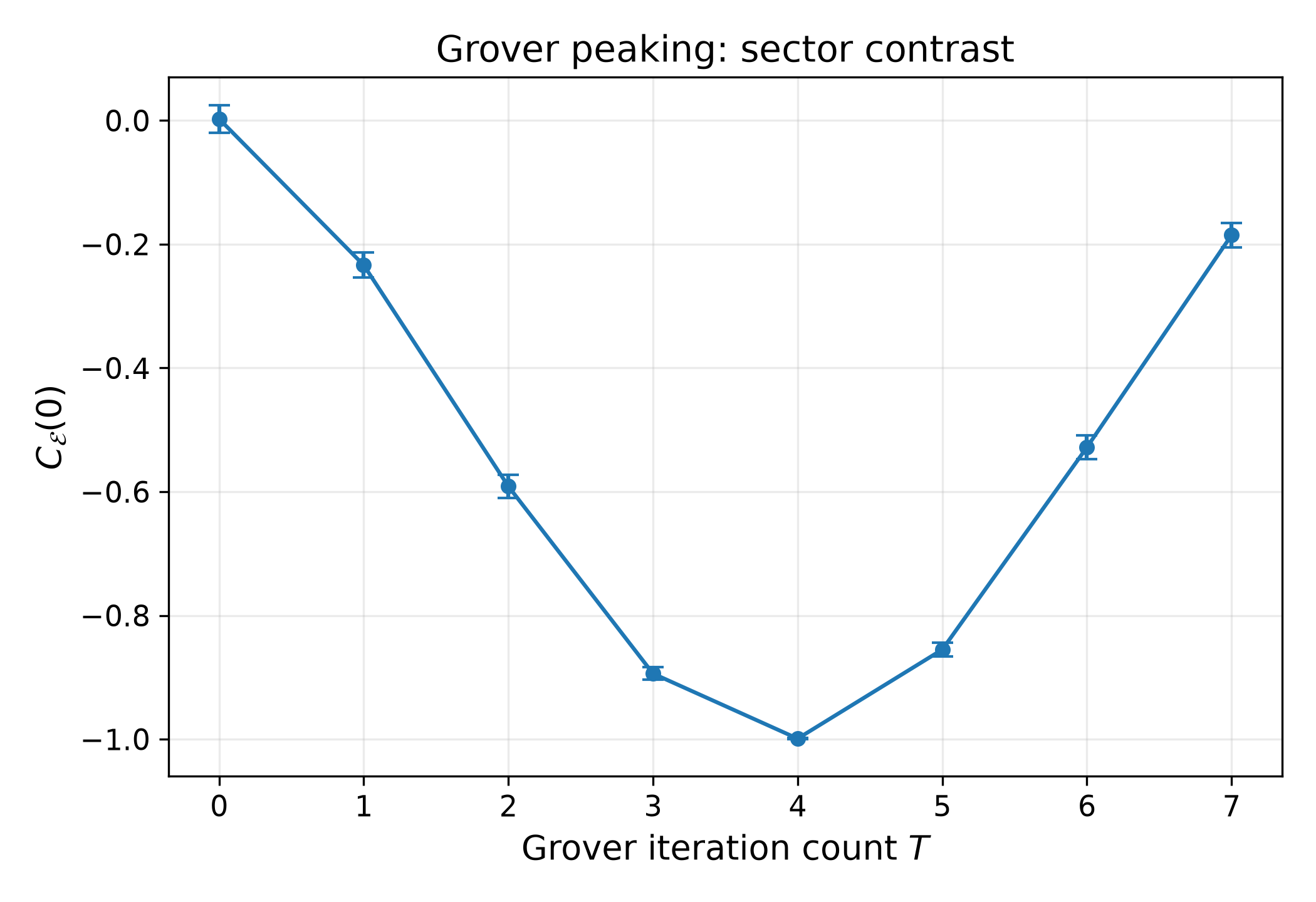}
    \end{minipage}
    \caption{Noiseless Grover-type parameter sweep with finite-shot sampling. Points show means over 100 independent 2048-shot samples of each fixed circuit; error bars indicate sample standard deviations.
    Left: sector mass $\pi_{\uparrow}$ versus Grover iteration count $T$.
    Right: sector contrast $C_E(0)$ versus $T$.
    Both quantities exhibit the non-monotonic, oscillatory dependence characteristic of amplitude amplification: the induced probability transfer and the resulting contrast peak near $T=4$ and then decrease at larger $T$.
    The Grover iteration count therefore provides discrete, oscillatory control and is sensitive to noise in hardware implementations.}
    \label{fig:grover_sim_sweep}
\end{figure*}

We first sweep the Grover iteration count $T$ \cite{grover1997needle}, which is the main control parameter of the oracle-based construction.
Starting from $H^{\otimes 10}|0\rangle^{\otimes 10}$, we mark
\[
\mathcal G=\{z:z_9=0,\ z_7=z_6=1,\ z_5=z_4=0\},
\]
so that $|\mathcal G|/2^{10}=1/32$ and every marked state has $a_z(0)b_z=-1$. We apply the phase oracle followed by diffusion for $T=0,\ldots,7$ iterations.
The weight visualizations introduced in Sec.~IV already show that increasing $T$ transfers probability mass into the marked subspace and produces a characteristic two-level structure in $q_z$.
Figure~\ref{fig:grover_sim_sweep} summarizes the resulting sector-resolved response at the level of the ensemble-dependent correlator.

At $T=0$, the initial mixing layer gives an exactly uniform probability distribution; finite-shot estimates fluctuate around $\pi_{\uparrow}=1/2$ and $C_E(0)=0$.
As $T$ increases, probability mass is transferred into the target sector and the contrast grows rapidly in magnitude:
$\pi_{\uparrow}$ rises from $0.5000\pm0.0108$ at $T=0$ to $0.9995\pm0.0005$ at $T=4$, while $C_E(0)$ evolves from $0.0027\pm0.0224$ to $-0.9991\pm0.0009$ over the same range.
Beyond that point, the sector mass and the magnitude of the contrast decrease again, reaching $\pi_{\uparrow}=0.5916\pm0.0110$ and $C_E(0)=-0.1852\pm0.0200$ at $T=7$.
The sweep thus displays the discrete, oscillatory control provided by $T$, as expected from amplitude amplification.

The close correspondence between the sector-mass and sector-contrast panels in Fig.~\ref{fig:grover_sim_sweep} is important.
In the Grover-type construction, the change in contrast is driven primarily by global probability transfer between the target and complement subspaces.
This same redistribution is also reflected in the empirical concentration diagnostics: for example, $M(64)$ grows from $0.1631\pm0.0048$ at $T=0$ to 1 in all 100 sampled repetitions at $T=4$, before decreasing again at larger $T$.
For the exactly uniform $T=0$ distribution, the exact concentration is $M(K)=K/2^{10}$, giving $M(64)=0.0625$. Comparison of the empirical value with this exact reference must account for the finite-shot ranking effect described above.
Thus, in the idealized setting, the redistribution that produces the characteristic peaked $q_z$ profile also produces the strongest sector imbalance.

These results clarify the role of the Grover-type construction in the remainder of the paper.
In principle, larger values of $T$ can produce very strong contrast, but the same mechanism requires repeated oracle and diffusion blocks, making the circuit increasingly fragile after transpilation.
The hardware section below therefore uses the smallest nontrivial setting, $T=1$, as a proof-of-principle operating point.

\subsection{Shallow peaking: depth-dependent concentration}
\label{subsec:shallow_sim}

We next consider the oracle-free shallow construction and sweep the number $d$ of complete rotation-and-entangling layers in a representative low-depth setting.
Starting from $|0\rangle^{\otimes 10}$, $U_{\mathrm{init}}$ applies Hadamards to qubits $6,4,1,8$, an $X$ gate to qubit 7, and $R_y(\pi/8)$, $R_y(\pi/10)$, and $R_y(\pi/12)$ to qubits $5,3,0$, respectively. Each shaping layer applies $R_y(\pi/12)$, $R_y(\pi/14)$, and $R_y(\pi/16)$ to qubits $6,4,1$, respectively, followed by CNOTs $6\to5$ and $4\to3$ in that order.
This preparation gives complete sector polarization, with $\pi_{\uparrow}=1$ and $\pi_{\downarrow}=0$ for all tested depths $d=0,1,2,3$.

For complete occupation, $\pi_{\uparrow}=1$, Eq.~\eqref{eq:contrast} reduces to $C_E(0)=m_{\uparrow}$; changes in intra-sector alignment can therefore change the contrast even at fixed occupation. In this particular sweep, $z_7=1$ and $z_9=0$ remain fixed, so $a_z(0)b_z=-1$ throughout the prepared support and $C_E(0)=-1$ at every depth. We examine how additional shaping layers redistribute probability \emph{within} this support: the coarse sector bias is already saturated, while the repeated rotations and sparse entangling gates reshape the intra-sector weights.

A product state has a factorized weight distribution $q_z = \prod_i q_i(z_i)$ and can exhibit both sector polarization and strong concentration. Combining shaping rotations with sparse entangling gates also allows correlations among the active bits, providing additional freedom to represent operator-dependent structure. The concentration metric $M(K)$ characterizes localization; correlations and alignment with the operator profile require separate diagnostics.

\begin{figure*}[t]
    \centering
    \begin{minipage}{0.49\textwidth}
        \centering
        \includegraphics[width=\linewidth]{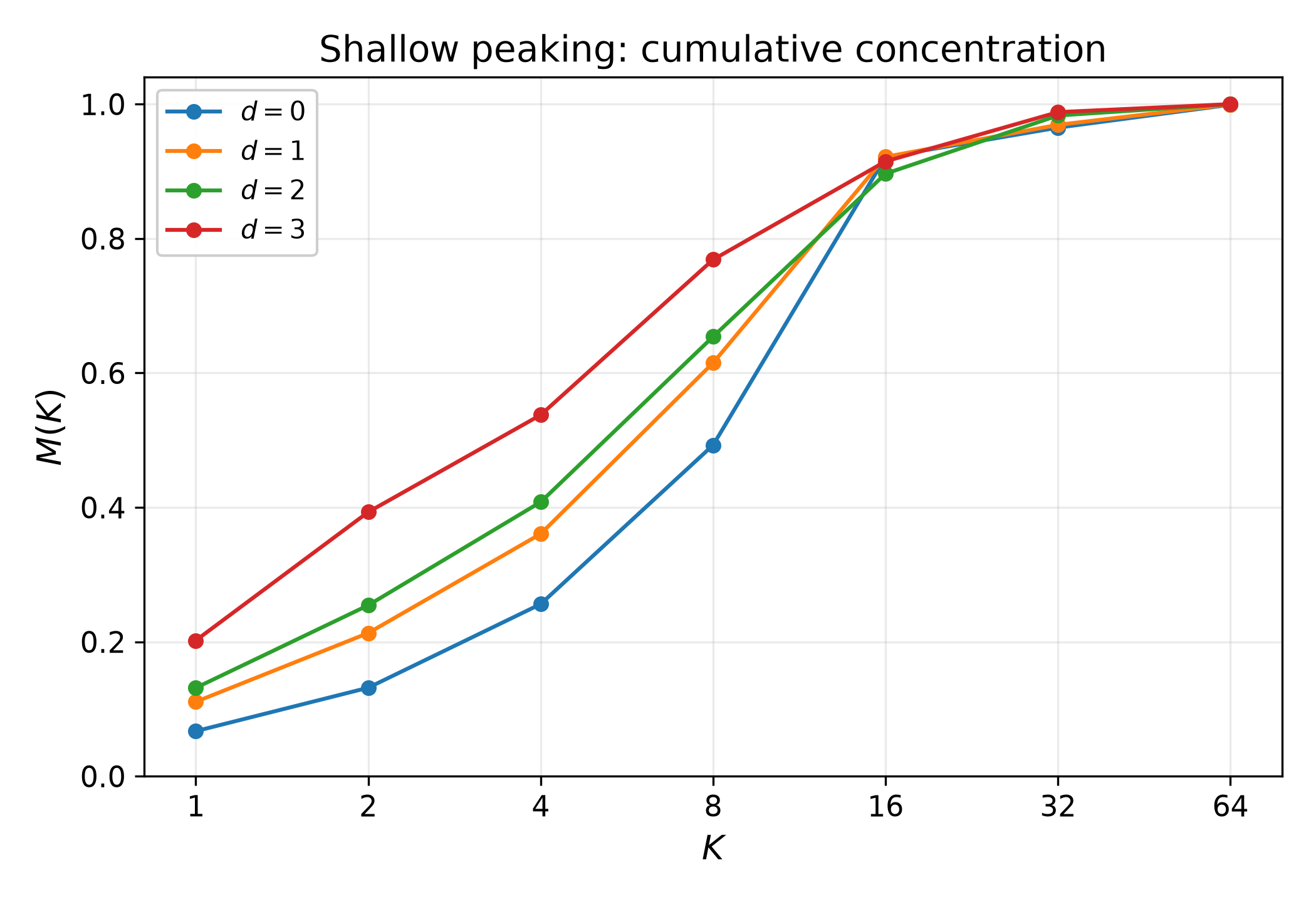}
    \end{minipage}
    \hfill
    \begin{minipage}{0.49\textwidth}
        \centering
        \includegraphics[width=\linewidth]{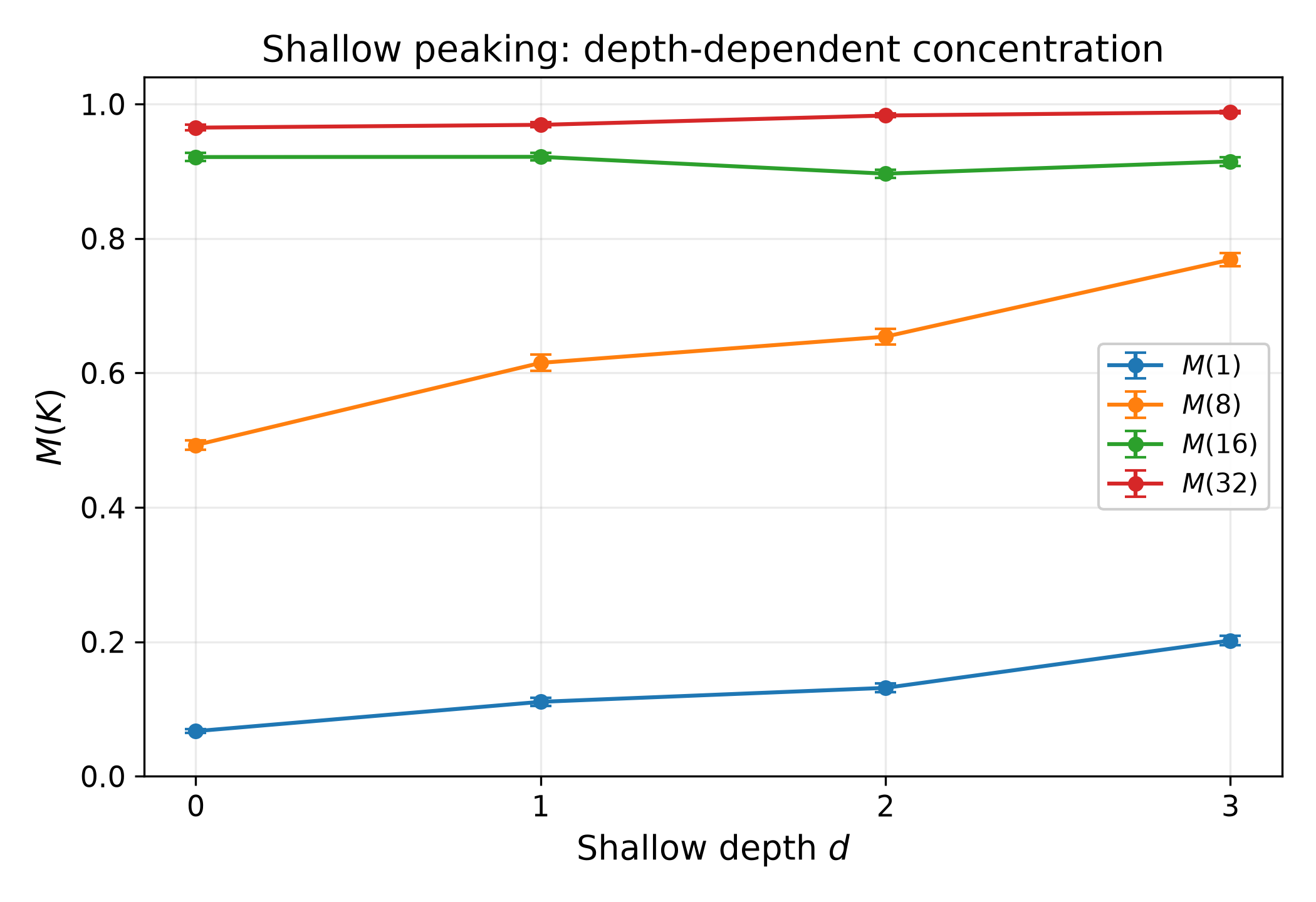}
    \end{minipage}
    \caption{Noiseless shallow-circuit parameter sweep with finite-shot sampling. Here $M(K)$ denotes the empirical top-$K$ mass $\widehat M(K)$ computed from sorted normalized measurement counts.
Left: mean cumulative top-$K$ mass for several shaping depths $d$.
Increasing depth shifts the curves upward at the plotted $K\le8$, indicating stronger localization onto fewer basis states within the selected support.
Right: means of selected $M(K)$ values versus $d$, with sample-standard-deviation error bars. Each depth uses 100 independent 2048-shot samples of a fixed circuit. In this parameter instance, $\pi_{\uparrow}=1$ and $C_E(0)=-1$ at every depth because $z_9=0$ and $z_7=1$ remain fixed.
The additional rotation-and-entangling layers reshape the weight distribution within the selected sector; $M(K)$ quantifies the resulting concentration. Sector occupation and intra-sector alignment jointly determine the contrast.}
    \label{fig:shallow_sim_sweep}
\end{figure*}

Figure~\ref{fig:shallow_sim_sweep} (left) shows the empirical cumulative top-$K$ mass $M(K)$ for several depths, calculated from sorted measurement frequencies as described above.
Increasing $d$ shifts the curve upward for the plotted $K\le8$, indicating that probability is being concentrated onto fewer basis states within the already selected sector.
The effect is substantial:
$M(1)$ increases from $0.0672\pm0.0028$ at $d=0$ to $0.2020\pm0.0066$ at $d=3$, and $M(8)$ from $0.4928\pm0.0074$ to $0.7687\pm0.0095$.
Higher-$K$ metrics are already closer to saturation; for example, $M(32)$ grows from $0.9650\pm0.0042$ to $0.9881\pm0.0020$ over the same range. The depth dependence need not be monotonic at every $K$, as the $M(16)$ series illustrates.
Figure~\ref{fig:shallow_sim_sweep}(right) presents the same point as a direct depth sweep of selected $M(K)$ values.

In this representative ideal instance, additional depth refines the intra-sector weight distribution while the sector occupation and contrast remain fixed. The resulting localization illustrates a low-depth operating regime relevant to NISQ hardware, with additional layers serving as a refinement of the prepared distribution.

\subsection{Distinct control mechanisms and relevance to hardware}
\label{subsec:sim_comparison}

The simulation sweeps identify distinct control roles for the two constructions.
Grover-type amplification changes both sector mass and sector contrast through a global, oscillatory redistribution of probability between marked and unmarked subspaces.
Shallow peaking, by contrast, achieves sector selection immediately in the present idealized instance and uses added depth mainly to sharpen localization within the selected support.

This distinction anticipates the hardware tradeoff below.
Grover-type peaking provides a clean mechanism-level benchmark with a clear ideal control parameter, but that control is inherently discrete and becomes fragile once repeated oracle and diffusion blocks must survive transpilation and noise.
The shallow construction offers less idealized global control, but it realizes its useful structure already at very low depth and is therefore naturally better matched to strict NISQ depth budgets.

\section{Demonstrations on IBM quantum hardware}
\label{sec:hardware}

\subsection{Demonstration strategy and common estimator}

The hardware experiments characterize how state preparation realizes sector populations and intra-sector alignment, and how these two features jointly determine the measured contrast. Following the noiseless characterization in Sec.~\ref{sec:simulation}, we apply a common measurement protocol and sector-resolved evaluation rule to the respective prepared ensembles.

For each preparation strategy, we measure sector populations and signed contributions in repeated hardware runs at a fixed circuit setting. Table~\ref{tab:hardware_summary} reports means and standard deviations over five independent runs per method to assess reproducibility under hardware fluctuations. Additional runs yielded the same qualitative conclusions. The uniform baseline serves as a structural reference and is omitted from the summary table.

The Grover-type construction serves as a small-scale benchmark, implemented on a 10-qubit instance where oracle and diffusion operations remain feasible after transpilation. The simulation results of Sec.~\ref{sec:simulation} show that larger values of $T$ can in principle produce stronger sector imbalance, but also that the control is oscillatory and requires repeated oracle and diffusion layers. Implementing the additional oracle and diffusion blocks increases the circuit resources required under hardware connectivity and depth constraints. We therefore use the smallest nontrivial setting, $T=1$, as the representative hardware benchmark.

The oracle-free shallow construction is evaluated in a larger 20-qubit setting, where strict depth constraints already limit the viability of Grover-type amplification but still permit low-depth biasing and sparse entanglement. In the corresponding simulation results of Sec.~\ref{sec:simulation}, sector selection is already achieved at minimal depth, so the relevant hardware question is whether this low-depth structure survives and remains observable at larger scale.

All hardware results reported here were executed on a single IBM Quantum backend, \texttt{ibm\_fez}, in order to avoid conflating device-to-device variation with the peaking effect itself, and all circuits were implemented and executed using the Qiskit software framework \cite{qiskit2024}.
Representative Grover results were obtained at 1024 and 2048 shots per circuit, with similar qualitative behavior; the summary statistics reported in Table~\ref{tab:hardware_summary} use the 2048-shot runs. For the representative shallow-construction demonstrations, we used a 20-qubit-scale hardware instance and evaluated the circuits at 4096 and 16384 shots per circuit, again observing similar qualitative trends; the summary statistics reported in Table~\ref{tab:hardware_summary} use the 16384-shot runs. All circuits were transpiled to the native gate set and connectivity of the chosen IBM backend and measured in the computational basis. For each measured bitstring $z$, the diagonal value $a_z=\langle z|A(0)|z\rangle$ is computed classically. The outcomes are then partitioned according to the sector label $z_k$, and the ensemble-dependent correlator $C_E(0)$ is evaluated using the sector-resolved quantities introduced in Sec.~IV. Here $q_z$ denotes the empirical probability estimated from measurement outcomes, corresponding to the basis-weight picture introduced in Sec.~III.

The hardware visualizations use the full-distribution cumulative sum $S_A(i)=\sum_{j=0}^{i}q_{z_j}a_{z_j}$, whose endpoint is $W_\uparrow+W_\downarrow$. For the uniform baseline, we show the signed contributions $q_z a_z$ together with $S_A(i)$. In Figs.~\ref{fig:cmp_grover} and \ref{fig:cmp_shallow}, we overlay hardware and simulation curves of $S_A(i)$ for the corresponding circuit template and parameter setting, using the same observable and integer-ordered computational basis. The red and blue line segments indicate the sign structure of $a_z$ in that ordering. These curves resolve the pattern of signed accumulation across the basis. The full correlator $C_E(0)=W_\uparrow-W_\downarrow$ is evaluated from the sector statistics reported in Table~\ref{tab:hardware_summary}. Each hardware curve illustrates an individual run, while the table summarizes the reported runs.

Comparisons with the corresponding noiseless circuit results identify changes in sector occupation, signed contributions, and cumulative profiles under hardware execution. Finite-shot variation and state-preparation, gate, and readout errors can all affect the recorded statistics; the present comparisons do not isolate their individual contributions. We therefore interpret the overlays through the observed pattern locations and accumulated magnitudes, together with the sector-resolved run statistics.

\begin{table*}[t]
    \centering
    \caption{Illustrative hardware demonstrations of the two peaked-state preparation strategies. Values are means $\pm$ standard deviations over five independent runs at a fixed circuit setting for each strategy.}
    \label{tab:hardware_summary}
    \begin{tabular}{lcccccccc}
        \hline
        Method & $n$ & shots & runs & $\pi_{\uparrow}$ & $\pi_{\downarrow}$ & $W_{\uparrow}$ & $W_{\downarrow}$ & $C_E(0)$ \\
        \hline
        Grover-type ($T = 1$)
        & 10 & 2048 & 5
        & $0.586 \pm 0.014$
        & $0.414 \pm 0.014$
        & $-0.168 \pm 0.011$
        & $0.011 \pm 0.006$
        & $-0.179 \pm 0.012$ \\
        Shallow peaking
        & 20 & 16384 & 5
        & $0.997 \pm 0.004$
        & $0.003 \pm 0.004$
        & $0.371 \pm 0.067$
        & $0.001 \pm 0.001$
        & $0.370 \pm 0.068$ \\
        \hline
    \end{tabular}
\end{table*}

\subsection{Uniform baseline: structural cancellation on hardware}
\label{subsec:uniform_hw}

As a reference, we first prepare the uniform superposition state
\begin{equation}
    |r\rangle = H^{\otimes n}|0\rangle^{\otimes n},
\end{equation}
which has an exactly uniform computational-basis distribution in the ideal circuit. For the diagonal Pauli products considered here, its exact correlator equals the uniform trace reference. Finite shots produce statistical deviations, while hardware imperfections can alter the measured distribution.

\begin{figure*}[t]
    \centering
    \includegraphics[width=\textwidth]{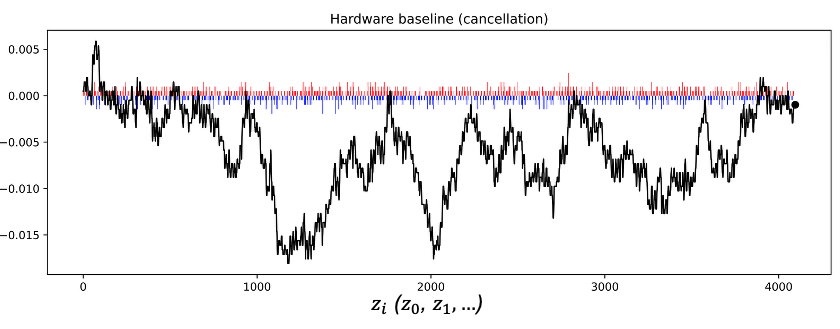}
    \caption{Hardware observation of structural cancellation under the uniform baseline.
    Bars show the signed basis contributions $q_z a_z$, while the black curve shows $S_A(i)=\sum_{j=0}^{i}q_{z_j}a_{z_j}$ as a function of the integer-ordered computational-basis index. Its endpoint is the empirical $A$ average $W_\uparrow+W_\downarrow$.
    Positive and negative contributions largely cancel in the displayed run. Finite-shot fluctuations and hardware imperfections can produce a residual relative to the ideal value $\mathrm{Tr}[A(0)]/2^n=0$.}
    \label{fig:uniform_hw}
\end{figure*}

Figure~\ref{fig:uniform_hw} displays the signed basis contributions $q_z a_z$ and $S_A(i)$ for a measured baseline instance. Positive and negative terms largely cancel. The ideal endpoint is $\mathrm{Tr}[A(0)]/2^n=0$ for the traceless Pauli observables considered here. The small measured residual $S_A(D-1)$ reflects finite sampling and hardware execution relative to this reference, separately from the state-to-state Haar fluctuations discussed in Sec.~\ref{subsec:trace_sampling}.

This baseline anchors the comparison of prepared ensembles. The peaked circuits are evaluated through changes in sector occupation, conditional alignment, and the resulting contrast relative to this known reference.

\subsection{Grover-type peaking as a small-scale hardware benchmark}
\label{subsec:grover_hw}

We next turn to the Grover-type construction, which serves as a structure-aligned small-scale benchmark for engineered peaking.
In the representative hardware demonstrations, we used a 10-qubit instance on \texttt{ibm\_fez} with a single Grover iteration ($T=1$).
The noiseless sweep of Sec.~\ref{sec:simulation} shows that stronger ideal contrast occurs at larger $T$, with additional oracle and diffusion blocks. We use the $T=1$ case as a reproducible nontrivial operating point for characterizing the implemented sector contributions at modest circuit depth.

\begin{figure*}[t]
    \centering
    \includegraphics[width=\textwidth]{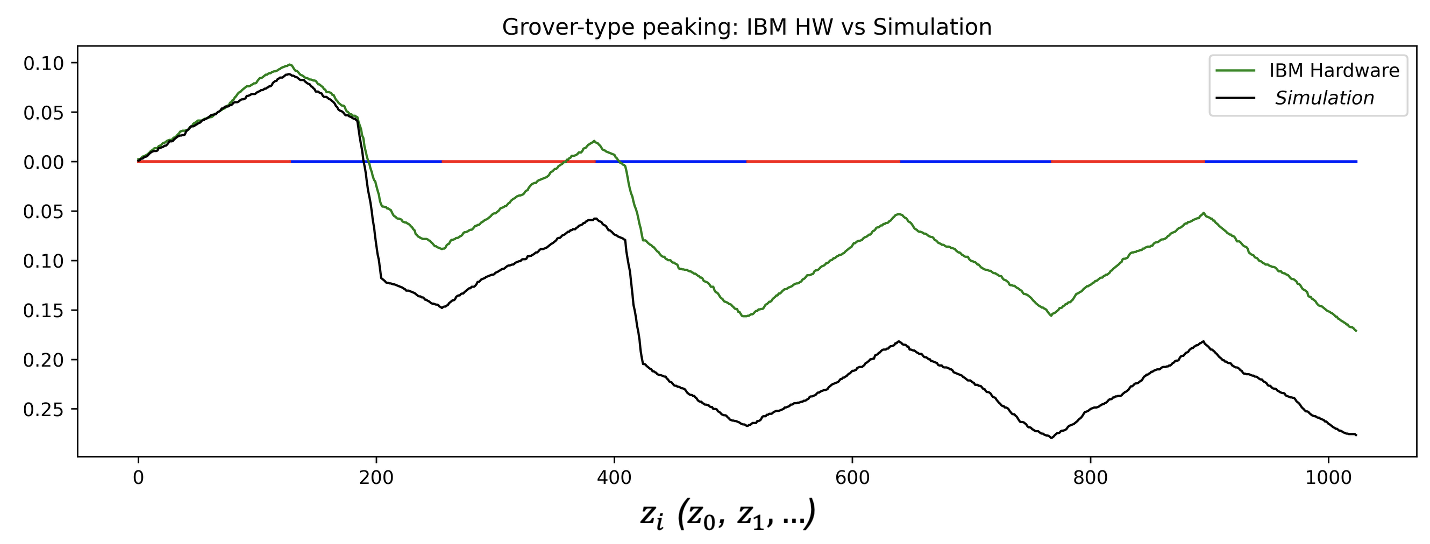}
    \caption{
        Comparison of Grover-type peaking on IBM quantum hardware and the corresponding noiseless simulation for a 10-qubit instance.
        The cumulative sum $S_A(i)=\sum_{j=0}^{i}q_{z_j}a_{z_j}$ is plotted as a function of the integer-ordered computational-basis index. Its endpoint is $W_\uparrow+W_\downarrow$; the full correlator is $W_\uparrow-W_\downarrow$.
        The red and blue line segments along the horizontal axis indicate basis regions associated with the sign structure of the diagonal operator profile $a_z = \langle z|A(0)|z\rangle$, which determines the target regions for peaking.
        The main jump locations agree closely between hardware and simulation, while the hardware curve has a smaller cumulative excursion and smoother local variations. The comparison resolves the retained support pattern and the change in its signed accumulation.
        }
    \label{fig:cmp_grover}
\end{figure*}

Figure~\ref{fig:cmp_grover} compares the cumulative signal $S_A(i)$ from a representative 10-qubit hardware run against the corresponding noiseless simulation. The red and blue line segments along the horizontal axis indicate the sign structure of the diagonal operator profile $a_z$ in the chosen basis ordering. The main jumps occur at nearly the same locations in the two curves, showing that the amplified-support structure targeted by the Grover step survives hardware execution. At the same time, the hardware curve exhibits a smaller overall negative excursion and smoother local variations than the noiseless reference, so the main support pattern remains visible while its signed accumulation is reduced.

The support-aligned jumps in $S_A(i)$ show where the prepared weighting produces signed accumulation in the $A$ average under hardware execution. The sector-resolved statistics below complete its connection to the full correlator.

The summary statistics in Table~\ref{tab:hardware_summary} show that this behavior is reproducible at the level of the sector-resolved estimator. Across five representative 2048-shot runs, the Grover-type construction yields $\pi_{\uparrow}=0.586\pm0.014$, $W_{\uparrow}=-0.168\pm0.011$, and $C_{\mathcal E}(0)=-0.179\pm0.012$, with a small residual opposite-sector contribution $W_{\downarrow}=0.011\pm0.006$. These values separate sector-mass transfer from intra-sector sign alignment: although $\pi_{\uparrow}$ indicates measurable probability transfer into the target sector, the smaller value of $|W_{\uparrow}|$ shows that the transferred weight is only partially aligned with the diagonal operator profile, leaving residual intra-sector cancellation. Relative to the ideal $T=1$ reference, the hardware signal is consistent with partial probability redistribution that persists under hardware execution and produces a resolved departure from the uniform-cancellation structure.

In this sense, the Grover data provide a proof of principle that amplitude amplification can be used as a distribution-engineering primitive for exposing operator-resolved structure that is hidden under uniform averaging, while also illustrating the distinction between probability transfer and full contrast optimization. Further oracle and diffusion blocks introduce additional gate and routing requirements. The reported 10-qubit instance establishes an implemented operating point for this form of ensemble control. Extending it to larger or deeper circuits requires assessing the corresponding implementation costs and errors.

\subsection{Shallow peaking in the 20-qubit-scale regime}
\label{subsec:shallow_hw}

We next examine the oracle-free shallow construction in the larger-qubit regime.
In the representative hardware demonstrations, we used a 20-qubit-scale instance on \texttt{ibm\_fez}.
Representative runs were obtained at 4096 and 16384 shots per circuit, and both shot counts produced the same qualitative peaking behavior and cumulative accumulation pattern.

\begin{figure*}[t]
    \centering
    \includegraphics[width=\textwidth]{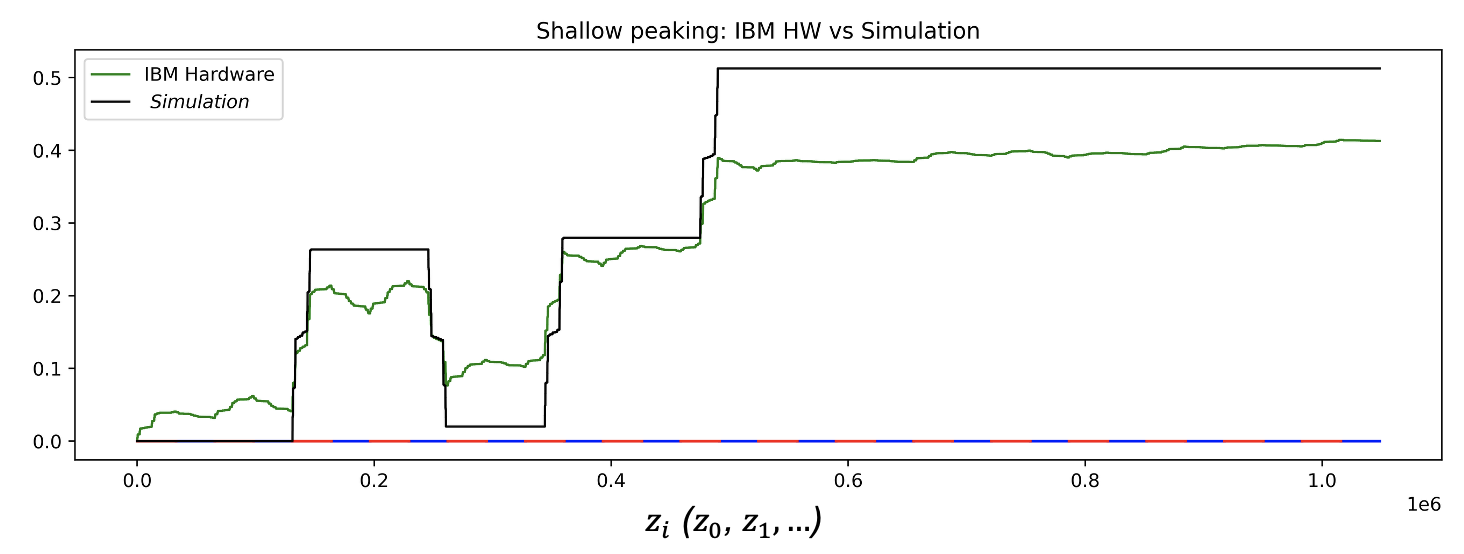}
    \caption{
        Comparison of oracle-free shallow peaking on IBM quantum hardware and the corresponding noiseless simulation in a 20-qubit setting.
        The cumulative sum $S_A(i)=\sum_{j=0}^{i}q_{z_j}a_{z_j}$ is shown as a function of the computational-basis index, which spans more than $10^6$ states. Its endpoint is $W_\uparrow+W_\downarrow$; the full correlator is $W_\uparrow-W_\downarrow$.
        The red and blue line segments indicate basis regions associated with the sign structure of the diagonal operator profile $a_z$, defining the operator-aligned regions targeted by the shallow construction.
        The hardware curve preserves the main step and plateau sequence of the ideal cumulative signal, indicating that operator-aligned localization survives at this scale.
        Relative to the noiseless curve, the measured transitions are rounded and the plateau heights are reduced.
        }
    \label{fig:cmp_shallow}
\end{figure*}

Figure~\ref{fig:cmp_shallow} compares the cumulative signal $S_A(i)$ from a representative 20-qubit-scale hardware run against the corresponding noiseless simulation, where the computational basis already contains more than $10^6$ states.
The red and blue line segments along the horizontal axis again indicate the sign structure of the diagonal operator profile $a_z$ in the chosen basis ordering.
The most striking feature is that the main step locations and plateau sequence of the ideal cumulative curve are preserved on hardware even at this scale.
The hardware curve has rounded steps, reduced plateau heights, and slow drift superimposed on the ideal structure.

The trajectory shows persistent net accumulation relative to the near-zero uniform baseline. At this scale, the overlay with the noiseless reference is more informative than plotting individual basis-resolved contributions, since those contributions are visually compressed by the size of the $2^{20}$-dimensional Hilbert space. The close correspondence of the step pattern nevertheless makes clear that the operator-aligned peaking survives in the larger-qubit regime.

The summary statistics in Table~\ref{tab:hardware_summary} show that this larger-scale behavior is also reproducible at the estimator level. Across five representative 16384-shot runs, the shallow construction exhibits near-complete single-sector concentration, $\pi_{\uparrow}=0.997\pm0.004$, together with a clearly resolved contrast, $C_E(0)=0.370\pm0.068$. The near-unity value of $\pi_{\uparrow}$ reflects the survival of the coarse sector-biasing component under hardware execution and should be interpreted primarily as a measure of baseline probability redistribution. By contrast, the sector-resolved contributions $W_{\uparrow}$ and $W_{\downarrow}$ determine how this redistributed weight aligns with the operator-dependent structure and therefore directly control the observed contrast. The observed contrast thus reflects both sector occupation and intra-sector alignment; the $S_A(i)$ curves show the cumulative pattern of the full-distribution contributions $q_z a_z$.

Figure~\ref{fig:cmp_shallow} shows the cumulative pattern of the $A$-weighted full distribution. The sign of the separately evaluated $C_E(0)$ is instance- and convention-dependent and need not match the representative idealized shallow sweep of Sec.~\ref{sec:simulation}.

The same qualitative $S_A(i)$ pattern is observed at both reported shot counts, while the repeated-run sector statistics characterize the reproducibility of the ensemble-dependent contrast.

\subsection{Practical comparison and implications}
\label{subsec:hardware_tradeoff}

For the Grover-type instance, the main support-determined jumps occur at similar locations in simulation and hardware, with reduced cumulative excursions on hardware. For the 20-qubit shallow instance, the principal step-and-plateau sequence remains visible with smoother transitions and lower plateaus.

The 12-qubit demonstrations in Ref.~\cite{Kim2025ADD} established hardware precedents for both preparation families and reported their projected signals. The present results characterize both sectors of the full correlator using five-run summaries for the 10-qubit Grover and 20-qubit shallow instances.

The sector diagnostics also explain what the observed contrast reveals about preparation. Using the rounded means in Table~\ref{tab:hardware_summary}, the ratios $\overline{W}_{\uparrow}/\overline{\pi}_{\uparrow}$ are approximately $-0.287$ for the Grover instance and $0.372$ for the shallow instance. These are ratios of the tabulated means, not estimates of the mean or uncertainty of the per-run ratios. For the shallow instance, near-complete sector occupation thus coexists with substantial cancellation within that sector. The Grover instance combines partial sector transfer with incomplete negative alignment. Measuring mass and signed contribution together distinguishes these features of the prepared distributions, which a final contrast or a concentration metric alone cannot specify.

The complementary-sector data also distinguish two reasons for a small correction to the projected signal. From the rounded means in Table~\ref{tab:hardware_summary}, $\overline{W}_{\downarrow}=0.011$ for Grover and $0.001$ for shallow, so $\overline{C}_{\mathcal E}(0)-\overline{W}_{\uparrow}$ is $-0.011$ and $-0.001$, respectively. In the Grover case, the small signed contribution coexists with an appreciable complementary-sector mass, $\overline{\pi}_{\downarrow}=0.414$, indicating cancellation within that occupied sector. In the shallow case, the complementary-sector mass itself is small, $\overline{\pi}_{\downarrow}=0.003$. The empirical version of Eq.~\eqref{eq:omitted_sector_bound}, averaged over runs, bounds the absolute mean correction by the corresponding mean mass. Resolving the signed contribution is therefore particularly informative for the Grover instance, whereas low occupation already bounds the shallow correction tightly. These are relations between the reported empirical quantities, not confidence bounds on the underlying expectation values.

Both hardware demonstrations realize the nonuniform weights through state preparation, without classical reweighting, postselection, or outcome filtering.

\section{Conclusion and Outlook}

Building on the peaked-state constructions and projected-sector diagnostics of Ref.~\cite{Kim2025ADD}, we have analyzed the full ensemble-dependent correlator using both sector contributions. Their sum gives the ensemble average and their difference gives the correlator. The worked example shows why identical selected-sector metrics need not determine the full correlator, while the elementary omission bound specifies when a selected-sector approximation is sufficient.

This interpretation is supported by the hardware data: across five representative runs, the 10-qubit Grover benchmark yielded a reproducible imbalance with $\pi_\uparrow = 0.586 \pm 0.014$ and $C_E(0) = -0.179 \pm 0.012$, while the 20-qubit shallow construction achieved near-complete single-sector concentration with $\pi_\uparrow = 0.997 \pm 0.004$ and retained an order-$10^{-1}$ contrast, $C_E(0) = 0.370 \pm 0.068$. Together with the signed contributions in Table~\ref{tab:hardware_summary}, these observations distinguish incomplete sector transfer from residual cancellation inside an occupied sector. In particular, the shallow example retains nearly complete sector occupation while its signed contribution remains well below that mass, demonstrating the information gained by resolving both quantities.

The present diagonal benchmarks focus on the regime near $t=0$ and measure $C_{\mathcal E}(t)$ for the prepared ensemble. They establish neither improved accuracy nor reduced shot cost for estimating the fixed trace $2^{-n}\mathrm{Tr}[A(t)B(0)]$. Within this scope, Grover-type peaking provides a mechanism-level benchmark, while shallow peaking supplies a low-depth route to preparing and measuring structured ensembles in the larger-qubit examples.
The resource analysis in Sec.~\ref{sec:peaked_circuits} makes the structural input explicit: known bit/parity rules admit constructive target specification and preparation, while more complex observables require a separate accounting of profile acquisition and target-design costs.
An immediate extension is to follow the sector contributions at finite time and determine when off-diagonal and phase-sensitive terms become significant. The multi-qubit diagonal analysis provides a starting point for studying more complex operator supports. Noise-aware circuit design, error mitigation, and compilation methods may help implement these extensions on NISQ hardware~\cite{temme2017errormitigation, noise1, noise2}.

With finite-time and non-diagonal extensions in place, ensemble engineering could also be explored for higher-order or scrambling-related observables, including OTOC-type correlators used to study quantum chaos and information spreading~\cite{PhysRevX.7.031011, swingle2018, landsman2019, xu2024scrambling,Manas-OTOC,Sophia-OTOC}. Such applications require specifying the prepared ensemble and the physical information sought from its correlator.

\section*{Acknowledgements}

We acknowledge funding from the US Department of Energy, the Office of Science through the Quantum Science Center (QSC), a National Quantum Information Science Research Center. M.S. would like to acknowledge the use of resources of the Oak Ridge Leadership Computing Facility at the Oak Ridge National Laboratory, which is supported by the Office of Science of the US Department of Energy under Contract No. DEAC05-00OR22725. This manuscript has in part been authored by UT-Battelle, LLC under Contract No. DE-AC05-00OR22725 with the U.S. Department of Energy.
The United States Government retains and the publisher, by accepting the article for publication, acknowledges that the US Government retains a nonexclusive, paid-up, irrevocable, worldwide license to publish or reproduce the published form of the manuscript, or allow others to do so, for US Government purposes. The Department of Energy will provide public access to these results of federally sponsored research in accordance with the DOE Public Access Plan (http://energy.gov/downloads/doe-publicaccess-plan).

\bibliography{apssamp}

\appendix

\section{IBM Quantum hardware-demonstration details}
\label{app:ibm_hardware_details}

This appendix summarizes the IBM Quantum hardware-demonstration details used in the manuscript.  The information below is provided to specify the backend, active physical qubits, measured-bit assignments, native-basis circuit summaries, calibration data, and raw-count processing used for the representative hardware demonstrations.

\subsection{Backend and calibration information}
\label{app:backend_calibration}

The representative hardware demonstrations were run on \texttt{ibm\_fez} on March 30, 2026.  The backend configuration queried through Qiskit reported a 156-qubit device with backend version 1.3.37 and native basis gates
\begin{equation}
\{\mathrm{cz},\mathrm{id},\mathrm{rz},\mathrm{sx},\mathrm{x}\}.
\end{equation}
The IBM Quantum Platform identifies \texttt{ibm\_fez} as a Heron r2 fixed-frequency system with an approximate device-level qubit-frequency band of 5--7~GHz.  Per-qubit frequency values were not exposed by the backend-properties data available to us.  We therefore report the platform-level frequency band together with the coherence, readout, and gate-error data returned by the backend-properties snapshots for the demonstration date.  The calibration data below are reported for one representative hardware-demonstration run from each circuit family.

\subsection{Representative 20-qubit shallow-construction demonstration}
\label{app:shallow_20q}

The representative 20-qubit shallow-construction demonstration used 20 active physical qubits on \texttt{ibm\_fez}.  The measured-bit to physical-qubit assignment, which also specifies the active physical-qubit set, is listed in Table~\ref{tab:shallow_mapping}.  The bit index \(z_i\) follows the classical register index \(c[i]\) used in the raw count dictionary.

\begin{table}[h]
\caption{Measured-bit to physical-qubit assignment for the representative 20-qubit shallow-construction hardware demonstration on \texttt{ibm\_fez}.}
\label{tab:shallow_mapping}
\begin{ruledtabular}
\begin{tabular}{cc@{\qquad}cc}
Measured bit & Physical qubit & Measured bit & Physical qubit \\
\hline
\(z_0\) / \(c[0]\)   & 110 & \(z_{10}\) / \(c[10]\) & 50  \\
\(z_1\) / \(c[1]\)   & 154 & \(z_{11}\) / \(c[11]\) & 84  \\
\(z_2\) / \(c[2]\)   & 14  & \(z_{12}\) / \(c[12]\) & 143 \\
\(z_3\) / \(c[3]\)   & 132 & \(z_{13}\) / \(c[13]\) & 136 \\
\(z_4\) / \(c[4]\)   & 9   & \(z_{14}\) / \(c[14]\) & 23  \\
\(z_5\) / \(c[5]\)   & 8   & \(z_{15}\) / \(c[15]\) & 24  \\
\(z_6\) / \(c[6]\)   & 134 & \(z_{16}\) / \(c[16]\) & 16  \\
\(z_7\) / \(c[7]\)   & 123 & \(z_{17}\) / \(c[17]\) & 22  \\
\(z_8\) / \(c[8]\)   & 124 & \(z_{18}\) / \(c[18]\) & 3   \\
\(z_9\) / \(c[9]\)   & 120 & \(z_{19}\) / \(c[19]\) & 21  \\
\end{tabular}
\end{ruledtabular}
\end{table}

After transpilation to the native basis of \texttt{ibm\_fez}, the representative shallow-construction circuit used 108 CZ gates and 20 measured qubits.  The physical CZ couplings used by the circuit were
\begin{equation}
\begin{split}
&(3,16),\ (8,9),\ (16,23),\ (21,22),\ (22,23),\ (23,24),\\
&(123,124),\ (136,143).
\end{split}
\end{equation}

\subsection{Representative 10-qubit Grover-type demonstration}
\label{app:grover_10q}

The representative 10-qubit Grover-type construction was built from an initial Hadamard layer, a marked-support oracle, and a diffusion step.  Before backend-specific transpilation, the logical circuit used a 10-qubit quantum register and a 10-bit classical register.  The hardware-submitted circuit was transpiled to the native IBM basis.

The representative hardware-submitted Grover-type circuit used 10 active physical qubits on \texttt{ibm\_fez}, all of which were measured.  The measured-bit to physical-qubit assignment, which also specifies the active physical-qubit set, is listed in Table~\ref{tab:grover_mapping}.

\begin{table}[h]
\caption{Measured-bit to physical-qubit assignment for the representative 10-qubit Grover-type hardware demonstration on \texttt{ibm\_fez}.}
\label{tab:grover_mapping}
\begin{ruledtabular}
\begin{tabular}{cc@{\qquad}cc}
Measured bit & Physical qubit & Measured bit & Physical qubit \\
\hline
\(z_0\) / \(\mathrm{meas}[0]\) & 111 & \(z_5\) / \(\mathrm{meas}[5]\) & 98  \\
\(z_1\) / \(\mathrm{meas}[1]\) & 110 & \(z_6\) / \(\mathrm{meas}[6]\) & 112 \\
\(z_2\) / \(\mathrm{meas}[2]\) & 2   & \(z_7\) / \(\mathrm{meas}[7]\) & 109 \\
\(z_3\) / \(\mathrm{meas}[3]\) & 134 & \(z_8\) / \(\mathrm{meas}[8]\) & 118 \\
\(z_4\) / \(\mathrm{meas}[4]\) & 50  & \(z_9\) / \(\mathrm{meas}[9]\) & 108 \\
\end{tabular}
\end{ruledtabular}
\end{table}

\begin{table*}[!t]
\caption{Calibration ranges for the active physical qubits and physical CZ couplings used in the representative hardware demonstrations.  The \(\sqrt{X}\)-error column reports the backend-property error for the native \(\sqrt{X}\) gate.}
\label{tab:calibration_summary}
\begin{ruledtabular}
\begin{tabular}{lcccccc}
Circuit & Active qubits & \(T_1\) (\(\mu\mathrm{s}\)) & \(T_2\) (\(\mu\mathrm{s}\)) & Readout error & \(\sqrt{X}\) error & CZ error \\
\hline
Shallow construction & 20 & 93.8--283.7 & 22.7--260.9 & 0.00427--0.0385 & \(1.53\times10^{-4}\)--\(1.38\times10^{-3}\) & 0.00176--0.00865 \\
Grover type          & 10 & 107.3--292.1 & 20.4--181.7 & 0.00464--0.0284 & \(1.16\times10^{-4}\)--\(3.03\times10^{-4}\) & 0.00196--0.00508 \\
\end{tabular}
\end{ruledtabular}
\end{table*}

After transpilation to the native basis of \texttt{ibm\_fez}, the representative Grover-type circuit used 129 CZ gates and 10 measured qubits.  The physical CZ couplings used by the circuit were
\begin{equation}
\begin{split}
&(98,111),\ (108,109),\ (109,110),\ (109,118), \\
&(110,111),\ (111,112).
\end{split}
\end{equation}

\subsection{Calibration summary}
\label{app:calibration_summary}

Table~\ref{tab:calibration_summary} summarizes the calibration data returned by the backend-properties snapshots for the active qubits and physical CZ couplings used by the representative demonstrations.  The entries are ranges over the active qubits or active couplings for each circuit family.  The readout length was 1560~ns for all active qubits, the single-qubit \(\sqrt{X}\) gate length was 24~ns for all active qubits, and the CZ gate duration was 68~ns for all used CZ couplings.


Combining the calibration ranges reported for the two representative runs in Table~\ref{tab:calibration_summary} gives \(T_1=93.8\)--292.1~\(\mu\mathrm{s}\), \(T_2=20.4\)--260.9~\(\mu\mathrm{s}\), readout errors of 0.00427--0.0385, and \(\sqrt{X}\) errors of \(1.16\times10^{-4}\)--\(1.38\times10^{-3}\). The two circuits together used 27 distinct active physical qubits and 14 distinct physical CZ couplings. The combined CZ-error range reported in the table is 0.00176--0.00865.

\subsection{Raw-count processing}
\label{app:raw_count_processing}

For each hardware demonstration, the IBM sampler returned a raw measurement-count dictionary.  The empirical hardware distribution was computed as
\begin{equation}
P_{\rm hw}(z)=\frac{N(z)}{N_{\rm shots}},
\end{equation}
where \(N(z)\) is the number of shots returning the measured bitstring \(z\).  The bit index \(z_i\) is defined by the classical register index shown in Tables~\ref{tab:shallow_mapping} and~\ref{tab:grover_mapping}.  When using Qiskit count strings, the string ordering was converted consistently to this classical-register convention before evaluating the diagnostics.

No outcome filtering, postselection, or classical reweighting was applied.  The sector-resolved and cancellation-related quantities reported in the main text were computed directly from the empirical count distribution.  For a generic diagnostic function \(F(z)\), the corresponding hardware estimate was evaluated as
\begin{equation}
\langle F\rangle_{\rm hw}=\sum_z P_{\rm hw}(z)F(z).
\end{equation}

\end{document}